\documentclass[a4paper,11pt]{article}\usepackage[]{graphicx}\usepackage[]{color}\usepackage{url}

\definecolor{fgcolor}{rgb}{0.345, 0.345, 0.345}
\definecolor{shadecolor}{rgb}{.97, .97, .97}
\definecolor{messagecolor}{rgb}{0, 0, 0}
\definecolor{warningcolor}{rgb}{1, 0, 1}
\definecolor{errorcolor}{rgb}{1, 0, 0}

\usepackage{graphicx}
\usepackage{mathptmx}      
\usepackage[T1]{fontenc}
\usepackage[utf8]{inputenc}
\usepackage[english]{babel}
\usepackage{subcaption}
\usepackage{amsmath} 

\usepackage{amsthm} 
\usepackage{amssymb} 
\usepackage{mathtools}
\usepackage{comment}
\usepackage{tikz}
\usepackage{multirow} 
\usepackage[abs]{overpic}
\usepackage{rotating}

\newcommand\RR{\mathbb{R}}

\newcommand\E{\text{E}}

\newcommand\Sum{\sum\limits}

\newtheorem{definicio*}{Definition}

\newtheorem*{lema-n}{Lemma}

\usepackage{fancyhdr}
\fancyheadoffset{30pt} 
\fancyfootoffset{30pt} 

\def\({\left(}
\def\){\right)}
\def\[{\left[}
\def\]{\right]}
\def\zij{z_{ij}}

\def\Sum{\sum\limits}
\def\Prod{\prod\limits}
\def\hatf{\hat{f}}

\IfFileExists{upquote.sty}{\usepackage{upquote}}{}
\begin{document}

\title{Relating Diversity and Human Appropriation \\ from Land Cover Data
  \thanks{This work has been partially
    supported by grant number  
    UNAB10-4E-378, co-funded by the European Regional Development Fund (ERDF); 
    grant number HAR2015-69620-C2-1-P funded by MINECO, and 
    the International Partnership Grant SSHRC- 895-2011-1020, 
    funded by the Social Sciences and Humanities Research Council of Canada.
    }
}

\author{Carme Font \\
        Department of Mathematics \\
        Universitat Aut\`onoma de Barcelona \\  
        08193 Bellaterra, Catalonia \\  
        \url{carmefont@mat.uab.cat} 
       \and
        Mercè Farré \\
        Department of Mathematics \\
        Universitat Aut\`onoma de Barcelona \\  
        08193 Bellaterra, Catalonia \\  
        \url{farre@mat.uab.cat} 
                           \and
        Aureli Alabert \\
        Department of Mathematics \\
        Universitat Autònoma de Barcelona \\  
        08193 Bellaterra, Catalonia \\  
        \url{Aureli.Alabert@uab.cat}  %
}

\date{\today} 
\maketitle
\thispagestyle{empty}
{\small
\begin{abstract}
We present a method to describe the relation between indicators of landscape diversity and the 
human appropriation of the net primary production in a given region.
These quantities  
are viewed as functions of the vector of proportions of the different land covers, which is 
in turn treated as a random vector whose values depend on the particular small terrain cell 
that is observed. 

We illustrate the method assuming first that the vector of proportions follows a uniform
distribution on the simplex. We then consider as starting point a raw dataset of
observed proportions for each cell, for which we must first obtain an estimate of its 
theoretical probability distribution, 
and secondly generate a sample of large size from it. 
We apply this procedure to real historical data of the Ma\-llor\-ca Island in three different moments
of time.

Our main goal is to compute the mean value of the landscape diversity as a function of
the level of human appropriation. This function is related to the so-called Energy-Species
hypothesis and to the Intermediate Disturbance Hypothesis.



\end{abstract}
}

\tableofcontents
\section{Introduction}
The \emph{Net Primary Production} (NPP) is the net amount of solar 
energy converted to plant organic matter through photosynthesis.
The \emph{Human Appropriation of Net Primary Production} (HANPP) 
is an indicator of the alterations produced by human activity on 
the NPP (see, for instance, 
\cite{Vitousek01061986}, \cite{Haberl2004213}, \cite{Haberl31072007}).
 These alterations include the
degradation of the environment (which leads to differentiate 
between the potential NPP and the actual NPP)
and the harvesting of photosynthetic products, which further 
reduces the actual NPP to a quantity sometimes
denoted $\text{NPP}_\text{t}$. 
Thus, $\text{HANPP} = \text{NPP}_{\text{pot}} - \text{NPP}_\text{t}$.

It is customary to measure the human appropriation as a percentage
of the potential primary production: 
$\text{HANPP\%} = 100\times\text{HANPP/NPP}_\text{pot}$.
One way to approximate the HANPP\% of a given area is to assign a coefficient 
$w_i$ to each of the $n$ different land uses present in the area and compute 
the weighted average 
\begin{equation}\label{eq:HANPP}
  \text{HANPP\%}=\sum_{i=1}^n w_ip_i
  \ ,
\end{equation}
  where $p_i$ are the proportions of land devoted to each use. The weights $w_i$ indicate
  the percentage of human appropriation for each specific land use. We will speak more generally
  of \emph{land covers} (forest, wetlands, crop, etc.). 
\bigskip

Ideally, we would like to relate human appropriation with some measure of the 
biodiversity in a given agro-ecosystem, in order to assess how human activity affects other species. 
                                              
There are several indices aimed at measuring biodiversity.
The most popular one is the \emph{Shannon index}, defined by the entropy formula 
\begin{equation*}
  H=-\sum_{k=1}^s q_k\log q_k
  \ ,
\end{equation*}               
  where $q_i,\ i=1,\dots,s$ is the proportion of each of the $s$ species of a 
  certain group which are present in a certain ecosystem.
  
 The Shannon index is sensitive both to the species richness and to its evenness in the 
 following precise sense: If, for some $j$, $0\le q_j<q_k$ holds for all $k\neq j$, then a 
 small increase in $q_j$ without increasing any of the other proportions results in 
 an increase of $H$. The base of the logarithm is arbitrary; if we take base $s$, then
 $0\le H\le 1$. 
   
 The \emph{Simpson diversity index}

$  
1-\sum_{k=1}^s q_k^2$            
  also increases with species richness and evenness, in the same sense above, whereas 
  the less used \emph{Berger--Parker index}
  $(\max q_i)^{-1}$ is only sensitive to the proportion
  of the most populated species.
  These indices, and some others, can be seen as particular cases of a family of measures
  (see \cite{leinster12}).
                                                                
  When we say ``number of species'' we are of course talking of a given taxonomic group
  of living organisms (such as birds, butterflies, trees, insects, mammals, 
  herbivores, carnivores, primary producers, \dots),
  possibly grouping together similar species.
  Obtaining an actual biodiversity index in a given region by direct observation
  and sampling is very difficult \cite{colwell09}.

Suppose anyway that 
we have a good estimate of the proportion of species 
in each particular land cover. Assume that we have $n$ different land covers 
coexisting in a given area in proportions 
$p_i,\ i=1,\dots,n$. Let 
$s$ be the total number of species in the area, and $q_{ik}$ the proportion of 
species $k$ in cover $i$. Then the Shannon index of the area is 
\begin{equation}\label{eq:biolandShannon}
  - \sum_{i=1}^{n} \sum_{k=1}^{s} q_{ik} p_i \log(q_{ik} p_i)
  \ .
 \end{equation}   
                
 In this formula we are assuming that species living in different covers 
 are considered different, thus in fact it combines bio-- and land-cover--diversity. 
 Eventually, the proximity of some covers may produce the appearance of new species 
 that are not present when the covers are not close (see again \cite{colwell09}).  

\bigskip

According to the so-called \emph{species-energy hypothesis} 
(see e.g. the survey by \cite{Currie2004} on this and 
other hypotheses, and the references therein),
the richness of species is monotonically increasing as a function of the
available energy in the system. This would explain, for example, the richness
gradient from the poles to the tropics, as the energy provided by the sun is
greater at lower latitudes. 

At geographical (large) scales, 
it has been suggested that this is true 
through all energy levels, 
although there is still little empirical evidence in this generality.
At local
scales this is not at all clear, and \cite{gaston2000}, among others, 
writes that ``there is a marked tendency for a general hump-shaped 
relationship between species richness and available energy''. 
In other words, that whereas when the available energy goes from 
low to moderate levels, richness indeed increase, from 
moderate to high levels the relation is reversed.

In terms of HANPP, which represents energy that humans take out
of the natural system, Gaston's remark amounts to say that 
biodiversity, as a function of HANPP, increases at the beginning,
peaks at a certain point, and then decrease again when HANPP is high.
The empirical work of \cite{Haberl2004213},
who measured the number of species of 9 groups (plant and animal)
on 38 small Austrian regions of similar characteristics, confirms that above 
40-50\% of total possible HANPP, species richness
indeed decreases, but there are no data below these percentages. 
The authors adjust a linear decreasing relationship, although
graphically the decrease seems to be more ``concave'' than linear 
in most cases. 

The possibility that low values of HANPP lead to diversity values
below the maximum seems to be related to the so-called
\emph{Intermediate Disturbance Hypotheses} (IDH), which states that 
moderate disturbances or fluctuations of any kind in an environment 
lead to more diversity than strong or weak disruptions.
It should be remarked that IDH is controversial, as it is the 
species-energy hypothesis. For instance, the recent review
article by \cite{fox2013} is clearly against. 
In any case, 
the intermediate disturbance in natural systems should be understood as
punctual interventions
or catastrophes, whereas in an agro-cultural system it is the result of the continuous
human intervention.

\bigskip

Numerous studies haven been published relating landscape heterogeneity 
with biodiversity. \cite{tews2004animal} contains a large review of articles 
on this subject; in most of them it is concluded that landscape diversity is positively 
correlated with species diversity.
With this fact in mind, and taking into account the difficulty to evaluate 
the biodiversity of a given area, 
we will use the Shannon index relative to land covers
\begin{equation}\label{eq:landShannon}
  H=- \sum_{i=1}^{n} p_i \log p_i
\end{equation}   
as our measure of `diversity'. 
The Simpson and Berger-Parker indices could be redefined in the same way,
replacing species proportions $q_k$ by land cover proportions $p_i$. 
Strictly speaking, however, (\ref{eq:landShannon}) is only 
an indicator of the degree 
of mosaic structure of a piece of land; 
\cite{Marull2015a} use a combination of $H$ and 
a so-called Ecological Connectivity Index to model biodiversity.

In this paper we try to relate human appropriation as defined by 
(\ref{eq:HANPP})
with the Shannon entropy index given by (\ref{eq:landShannon}).
For the sake of brevity, 
we will denote   
the HANPP\% measure 
of (\ref{eq:HANPP}) 
simply by $A$ in the formulae throughout the paper,
and we will speak of (human) \emph{appropriation}.

Both $H$ and $A$ are functions of the land proportions $p_i$ 
in a terrain cell, but we would like 
somehow to obtain a 
``function'' yielding $H$ from the appropriation alone.
Actually, this is not possible in a strict sense, since the same value of appropriation may
correspond to many values of entropy, and vice versa.
We propose the following setup: 

On a given terrain cell $\omega$, the different land covers may appear
in certain proportions $0\le p_i(\omega)\le 1$. 
Suppose we observe a big number of such cells, 
and we apply a fixed set of coefficients $w_1,\dots,w_n$ to 
all of them.

Then, cell $\omega$ has a certain appropriation 
$A(\omega)$ and a certain entropy $H(\omega)$. We may think that 
$\omega$ is a random parameter, so that $p_i(\omega)$ are random proportions
and also $A(\omega)$ and $H(\omega)$ are random. We aim at describing
the probability distribution of $H$ given a certain value $A(\omega)=a$ of the appropriation, for
every possible $a$.
It is therefore the probability distribution of $H$ which will be a function of $A$.
 
\bigskip 

\begin{figure}[h]
\centering
 \includegraphics[width=0.45\textwidth]{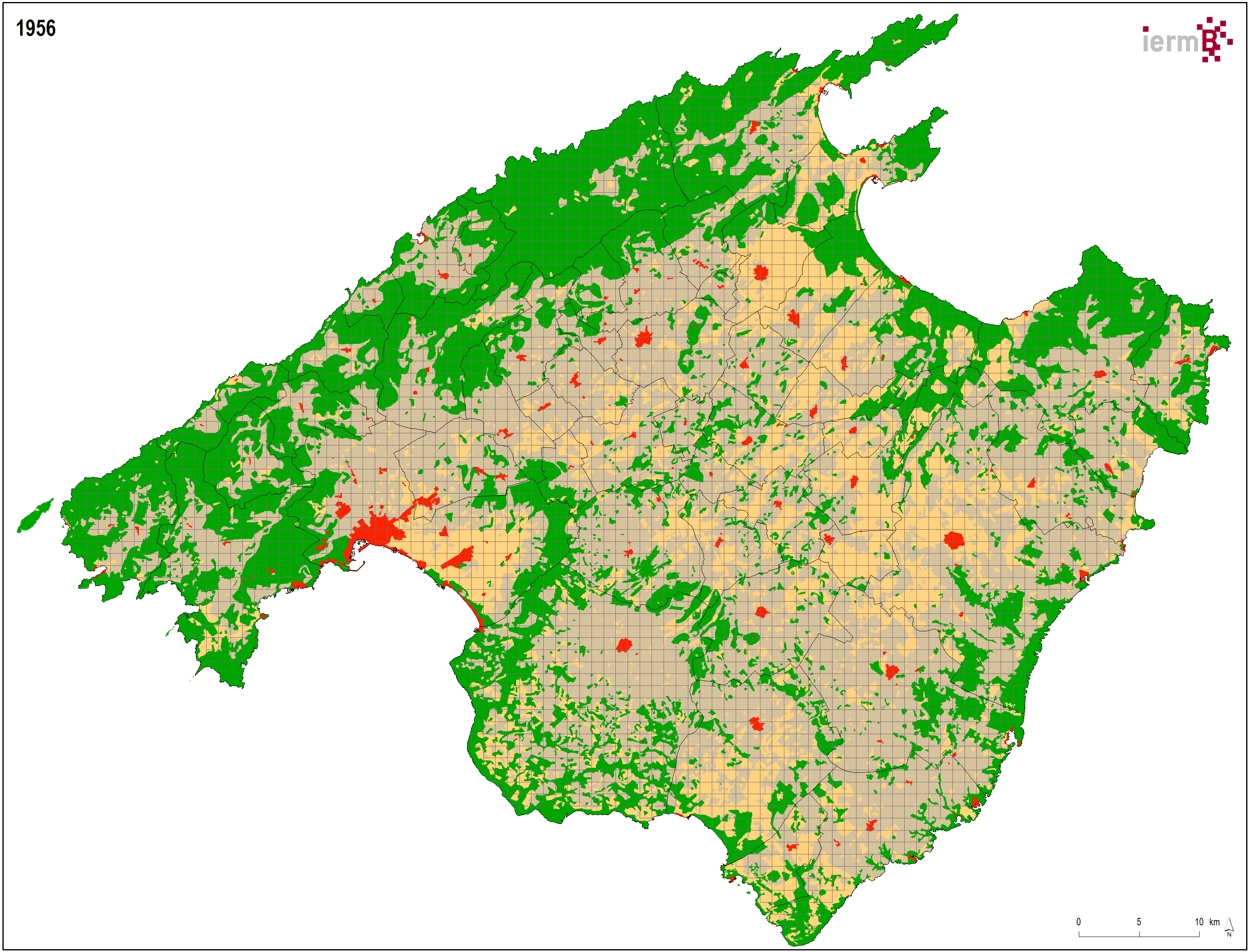}
 \includegraphics[width=0.45\textwidth]{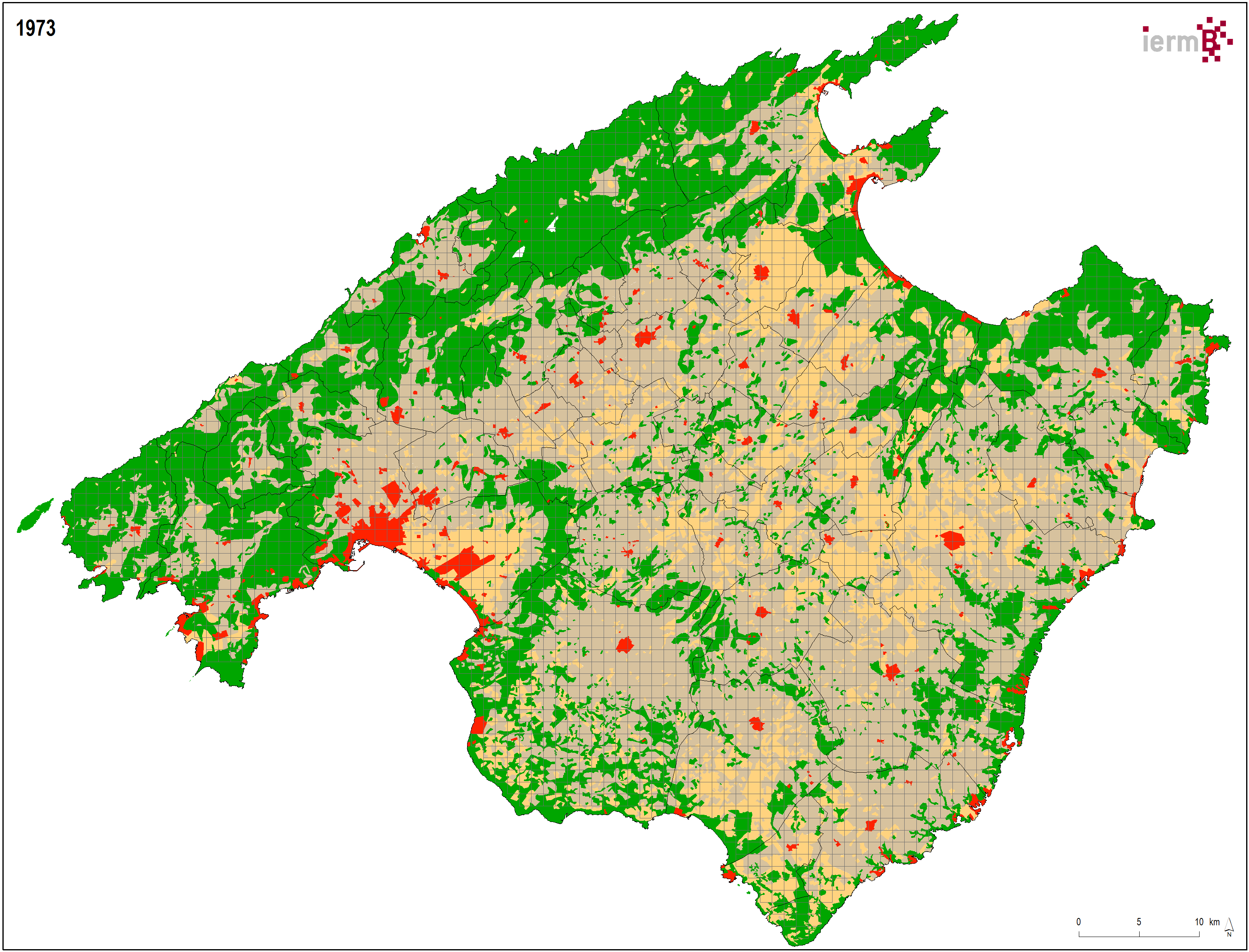}
 \begin{subfigure}{0.45\textwidth}\centering
 \includegraphics[width=1\textwidth]{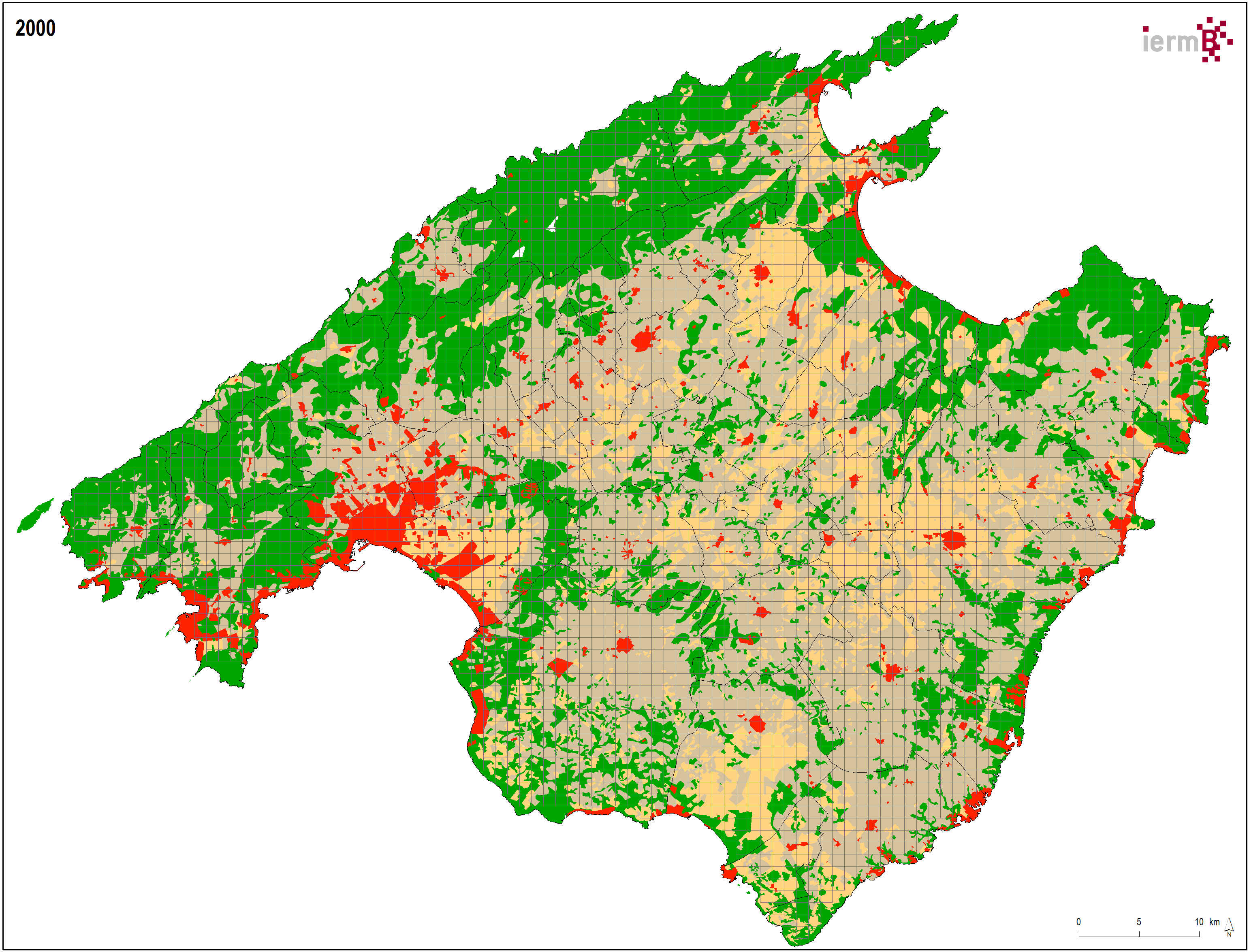}
 \end{subfigure}
 \begin{subfigure}{0.45\textwidth}\centering
\begin{tikzpicture}[x=1pt,y=1pt]
\definecolor{fillColor}{RGB}{255,255,255}
\path[use as bounding box,fill=fillColor,fill opacity=0.00] (10,110) rectangle (247.5,165);
\draw[help lines, step=10] (10.5,110) rectangle (247.3,165);
\begin{scope}
\path[clip] ( 49.20, 61.20) rectangle (263.88,167.61);
\definecolor{drawColor}{RGB}{0,0,0}

\node[text=drawColor,anchor=base,inner sep=0pt, outer sep=0pt, scale=  1.00] at (136.66,148.40) {\footnotesize Land Covers};
\definecolor{fillColor}{RGB}{0,166,0}

\path[fill=fillColor] ( 52.65,134.54) --
	( 61.65,134.54) --
	( 61.65,143.54) --
	( 52.65,143.54) --
	cycle;

\path[fill=fillColor] ( 56.63,134.54) --
	( 65.63,134.54) --
	( 65.63,143.54) --
	( 56.63,143.54) --
	cycle;

\path[fill=fillColor] ( 60.60,134.54) --
	( 69.60,134.54) --
	( 69.60,143.54) --
	( 60.60,143.54) --
	cycle;

\path[fill=fillColor] ( 64.58,134.54) --
	( 73.58,134.54) --
	( 73.58,143.54) --
	( 64.58,143.54) --
	cycle;

\node[text=drawColor,anchor=base,inner sep=0pt, outer sep=0pt, scale=  1.00] at (100.88,135.59) {\footnotesize semi-natural};
\definecolor{fillColor}{RGB}{255,211,127}

\path[fill=fillColor] ( 52.65,117.29) --
	( 61.65,117.29) --
	( 61.65,126.29) --
	( 52.65,126.29) --
	cycle;

\path[fill=fillColor] ( 56.63,117.29) --
	( 65.63,117.29) --
	( 65.63,126.29) --
	( 56.63,126.29) --
	cycle;

\path[fill=fillColor] ( 60.60,117.29) --
	( 69.60,117.29) --
	( 69.60,126.29) --
	( 60.60,126.29) --
	cycle;

\path[fill=fillColor] ( 64.58,117.29) --
	( 73.58,117.29) --
	( 73.58,126.29) --
	( 64.58,126.29) --
	cycle;

\node[text=drawColor,anchor=base,inner sep=0pt, outer sep=0pt, scale=  1.00] at ( 92.93,119.32) {\footnotesize cropland};
\definecolor{fillColor}{RGB}{215,194,158}

\path[fill=fillColor] (152.04,134.54) --
	(161.04,134.54) --
	(161.04,143.54) --
	(152.04,143.54) --
	cycle;

\path[fill=fillColor] (156.02,134.54) --
	(165.02,134.54) --
	(165.02,143.54) --
	(156.02,143.54) --
	cycle;

\path[fill=fillColor] (159.99,134.54) --
	(168.99,134.54) --
	(168.99,143.54) --
	(159.99,143.54) --
	cycle;

\path[fill=fillColor] (163.97,134.54) --
	(172.97,134.54) --
	(172.97,143.54) --
	(163.97,143.54) --
	cycle;

\node[text=drawColor,anchor=base,inner sep=0pt, outer sep=0pt, scale=  1.00] at (188.34,137.86) {\footnotesize groves};
\definecolor{fillColor}{RGB}{255,34,0}

\path[fill=fillColor] (152.04,117.29) --
	(161.04,117.29) --
	(161.04,126.29) --
	(152.04,126.29) --
	cycle;

\path[fill=fillColor] (156.02,117.29) --
	(165.02,117.29) --
	(165.02,126.29) --
	(156.02,126.29) --
	cycle;

\path[fill=fillColor] (159.99,117.29) --
	(168.99,117.29) --
	(168.99,126.29) --
	(159.99,126.29) --
	cycle;

\path[fill=fillColor] (163.97,117.29) --
	(172.97,117.29) --
	(172.97,126.29) --
	(163.97,126.29) --
	cycle;

\node[text=drawColor,anchor=base,inner sep=0pt, outer sep=0pt, scale=  1.00] at (186.36,118.35) {\footnotesize urban};
\end{scope}
\end{tikzpicture}
 \end{subfigure}
 \caption{Mallorca land cover maps at regional scale (1:50,000) for 1956, 1973 and 2000. Source: from \cite{GIST} in collaboration with the 
 Barcelona Institute of Regional and Metropolitan Studies.}
 \label{fig:maps}
\end{figure}
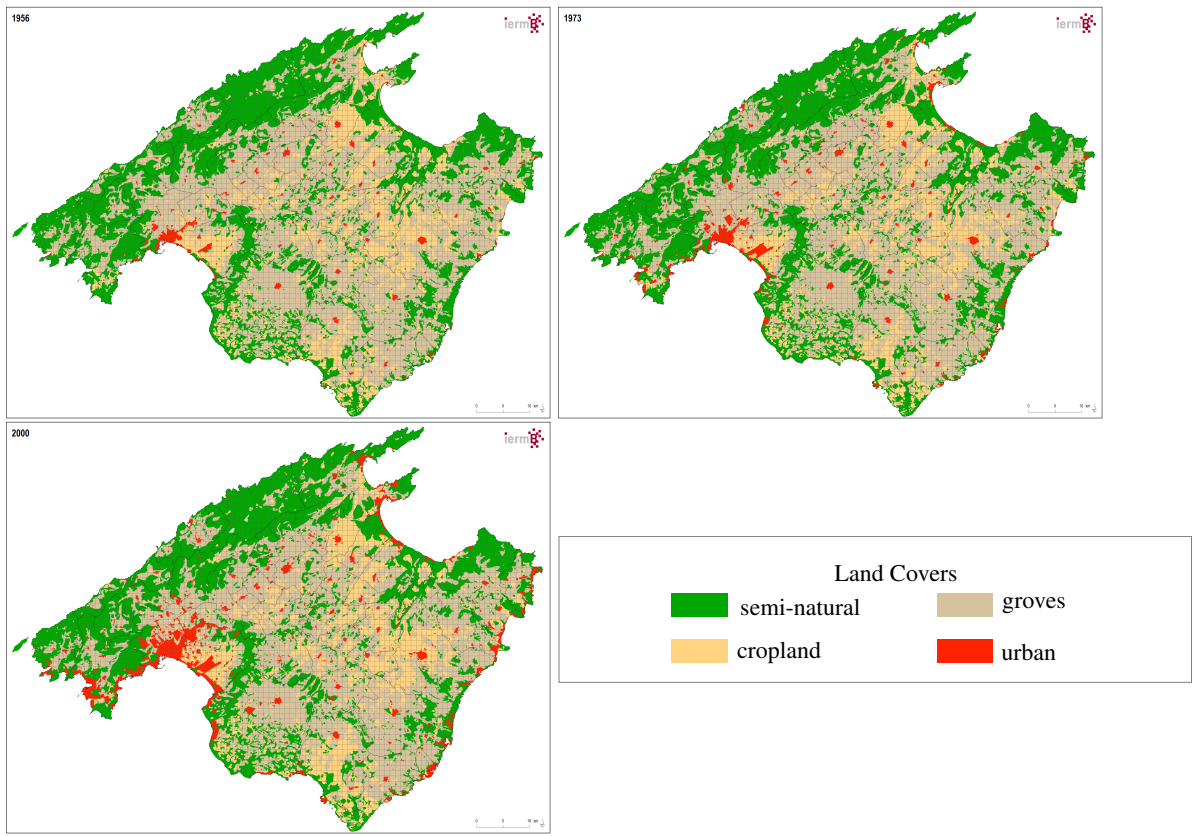

Our case study is \emph{Mallorca}, 
a Mediterranean island with a total area of 3,603 km$^2$ of calcareous origin. 
The mountain range of \emph{Serra de Tramuntana} runs parallel to the North coast and 
reaches 1,445 metres in
the highest peak. Between this range and the eastern mountains of \emph{Serres de Llevant}, a plain occupies
most of the island. Annual precipitation varies from 300 mm (in the South) to 1,800 mm (in the
North) with an average temperature of 16 ºC. 
We work with land cover data based on land cover
maps of Mallorca (Figure \ref{fig:maps}) obtained from \cite{GIST} 
 for three time periods (1956, 1973, 2000). These data comprises a total of 
 3360 cells of 
 size $1\times1$ km$^2$, once disregarded those with some part into the sea.

  We have grouped land covers into four categories, namely ‘semi-natural’, ‘croplands’, ‘groves’ 
  and ‘urban’. Semi-natural land covers include forest, scrub, prairie and bedrock, and wetlands. 
  Croplands include both dry and irrigated croplands. Groves are composed of 
  rain-fed arboricultural groves, irrigated groves and olive groves. 
  Urban land covers are both urban and industrial areas.

\bigskip

The methodology proposed here to relate the Shannon entropy $H$ 
with the appropriation $A$ can also be used with the other indices of 
diversity cited in this introduction, or with other functions of land cover proportions.
The human appropriation  
can be either a random variable whose distribution is determined by a given theoretical 
distribution of land-covers (case treated in Section \ref{sec:unifdist}), or a function of empirically
obtained data (developed in Section \ref{sec:realdata}, with our case study in mind).

In Section \ref{sec:unifdist}, we assume a simple uniform probability distribution of proportions
of land covers and 
\begin{enumerate}
  \item [a)]
  we show how to obtain (by simulation) the distribution of the entropy $H$, and we compute 
  (exactly) its expected value;
  \item [b)]
  we compute the distribution of the appropriation $A$ and its expectation, and
  \item [c)]
  we derive a formula for the conditional expectation of $H$ given any fixed value of the appropriation.
\end{enumerate}

In Section \ref{sec:realdata}, we estimate the conditional 
expectation of $H$ given $A$ using real sample data.
This involves estimating the probability distribution from 
which the data has been (ideally) originated, and produce a very large sample following
the estimated distribution. The process has some difficulties which are 
explained at the beginning of the section, and developed in several subsections.

Finally, some specific data-related details and the results of the case study are presented
in Section \ref{sec:results}.

\section{Uniform distribution of land covers}\label{sec:unifdist}
  Given a set of cells $\Omega$, and a set of $n+1$ possible land covers, 
  we have defined the appropriation and Shannon indices
  of each cell $\omega\in\Omega$, by 
\begin{equation}
\begin{split}
A(\omega)&=\Sum_{i=1}^{n+1} w_ip_i(\omega) 
\\
H(\omega)&=-\Sum_{i=1}^{n+1}p_i(\omega)\log_{n+1}p_i(\omega) \label{eq:AHomega}
\end{split}
\end{equation}
  where $p_i(\omega)$ is the proportion of cover $i$ in cell $\omega$, and we arbitrarily 
  take $n + 1$ as the base of the logarithm, so that the maximal value that $H$ can achieve is 
  normalised to 1. Working in dimension $n+1$ instead of $n$ simplifies the notation later.
  
  We study the relation between these two quantities by postulating some probability distribution 
  of the random vector $p(\omega)=\big(p_1(\omega),\dots, p_{n+1}(\omega)\big)$. Notice that 
  this vector takes values in the so-called \emph{standard $n$-simplex in $\RR^{n+1}$}, i.e. 
  the $n$-dimensional surface
\begin{equation*}
  \Delta=\left\{(p_1,\hdots,p_{n+1})\ |\ p_i\geq0,\ p_1+\cdots+p_{n+1}=1\right\} \ .
\end{equation*}
  We are thus working with \emph{compositional data} (see e.g. \cite{buccianti06}).

  In this section we will assume that $p$ follows the uniform distribution on the simplex.
  This assumption does not aim to represent any realistic situation; for instance, 
  it implies that all covers are actually present in some proportion in all cells. 
  But it is anyway the 
  usual modelling choice when no other information is present.
   
  The volume of the standard $n$-simplex is $\frac{\sqrt{n+1}}{n!}$, whence the density
  of the uniform distribution is given by 
\begin{equation*}
f(p_1,\hdots, p_{n+1})=
\begin{cases}
\frac{n!}{\sqrt{n+1}}&\text{ if }p\in \Delta\\
0&\text{ otherwise .}
\end{cases}
\end{equation*}

  The marginal distribution of the first $n$ coordinates is also uniform, on the projected
  simplex
\begin{equation*}
\Delta'=\left\{(p_1,\hdots, p_n)\ |\ p_i\geq0,\ p_1+\cdots+p_n\leq1\right\}\ ,
\end{equation*}
with the density
\begin{equation*}
f(p_1,\hdots, p_{n})=\begin{cases}
n!&\text{ if }p\in \Delta'\\
0&\text{ otherwise .}
\end{cases}
\end{equation*}

We can easily obtain the marginal density function of $p_i$ integrating $f$ with respect to $p_j$, $j\neq i$.
For $p_1$, 
\begin{align}\label{eq:p_marginal}
f(p_1) &=  n!\int_0^{1-p_1}\cdots \int_0^{1-\sum_{i=0}^{n-1}p_i} dp_{n}\cdots dp_{2}\nonumber\\ 
 &= n(1-p_1)^{n-1}\ ,
\end{align}
and by symmetry the same formula holds for all $p_i$.

  It is better to work in the projected simplex, since $f$ is then a true density
  with respect to Lebesgue measure in $\RR^n$, whereas on the standard simplex the
  support of the probability has zero measure as a subset of $\RR^{n+1}$.

In the next subsections we study the probability distribution of the random
variables $H$ and $A$, and the conditional expectation of $H$ given $A$. 
We will in general avoid to write explicitly
the random parameter $\omega$ from which the land covers depend.

\subsection{The distribution of $H$}
\begin{figure}
\begin{subfigure}{0.45\columnwidth}
  \centering
\begin{tikzpicture}[x=1pt,y=1pt]
\definecolor{fillColor}{RGB}{255,255,255}
\path[use as bounding box,fill=fillColor,fill opacity=0.00] (25,10) rectangle (228.94,228.94);
\begin{scope}
\path[clip] (  0.00,  0.00) rectangle (252.94,252.94);
\definecolor{drawColor}{RGB}{0,0,0}

\node[text=drawColor,rotate= 90.00,anchor=base,inner sep=0pt, outer sep=0pt, scale=  1.00] at ( -2.40,132.47) {Density};
\end{scope}
\begin{scope}
\path[clip] ( 36.00, 36.00) rectangle (228.94,228.94);
\definecolor{drawColor}{gray}{0.40}

\path[draw=drawColor,line width= 0.4pt,line join=round,line cap=round] ( 43.15, 43.15) rectangle ( 52.08, 43.44);

\path[draw=drawColor,line width= 0.4pt,line join=round,line cap=round] ( 52.08, 43.15) rectangle ( 61.01, 44.62);

\path[draw=drawColor,line width= 0.4pt,line join=round,line cap=round] ( 61.01, 43.15) rectangle ( 69.94, 46.11);

\path[draw=drawColor,line width= 0.4pt,line join=round,line cap=round] ( 69.94, 43.15) rectangle ( 78.88, 48.33);

\path[draw=drawColor,line width= 0.4pt,line join=round,line cap=round] ( 78.88, 43.15) rectangle ( 87.81, 51.34);

\path[draw=drawColor,line width= 0.4pt,line join=round,line cap=round] ( 87.81, 43.15) rectangle ( 96.74, 54.58);

\path[draw=drawColor,line width= 0.4pt,line join=round,line cap=round] ( 96.74, 43.15) rectangle (105.67, 59.17);

\path[draw=drawColor,line width= 0.4pt,line join=round,line cap=round] (105.67, 43.15) rectangle (114.61, 64.23);

\path[draw=drawColor,line width= 0.4pt,line join=round,line cap=round] (114.61, 43.15) rectangle (123.54, 70.72);

\path[draw=drawColor,line width= 0.4pt,line join=round,line cap=round] (123.54, 43.15) rectangle (132.47, 79.24);

\path[draw=drawColor,line width= 0.4pt,line join=round,line cap=round] (132.47, 43.15) rectangle (141.41, 90.61);

\path[draw=drawColor,line width= 0.4pt,line join=round,line cap=round] (141.41, 43.15) rectangle (150.34,105.55);

\path[draw=drawColor,line width= 0.4pt,line join=round,line cap=round] (150.34, 43.15) rectangle (159.27,133.13);

\path[draw=drawColor,line width= 0.4pt,line join=round,line cap=round] (159.27, 43.15) rectangle (168.20,158.71);

\path[draw=drawColor,line width= 0.4pt,line join=round,line cap=round] (168.20, 43.15) rectangle (177.14,172.98);

\path[draw=drawColor,line width= 0.4pt,line join=round,line cap=round] (177.14, 43.15) rectangle (186.07,185.08);

\path[draw=drawColor,line width= 0.4pt,line join=round,line cap=round] (186.07, 43.15) rectangle (195.00,194.44);

\path[draw=drawColor,line width= 0.4pt,line join=round,line cap=round] (195.00, 43.15) rectangle (203.93,204.36);

\path[draw=drawColor,line width= 0.4pt,line join=round,line cap=round] (203.93, 43.15) rectangle (212.87,213.00);

\path[draw=drawColor,line width= 0.4pt,line join=round,line cap=round] (212.87, 43.15) rectangle (221.80,221.80);
\definecolor{drawColor}{RGB}{255,0,0}

\path[draw=drawColor,line width= 0.4pt,line join=round,line cap=round] (221.80,227.09) --
	(219.99,225.08) --
	(218.19,223.09) --
	(216.39,221.12) --
	(214.58,219.17) --
	(212.78,217.24) --
	(210.97,215.34) --
	(209.17,213.45) --
	(207.36,211.59) --
	(205.56,209.75) --
	(203.75,207.92) --
	(201.95,206.12) --
	(200.14,204.33) --
	(198.34,202.57) --
	(196.53,200.82) --
	(194.73,199.06) --
	(192.93,197.29) --
	(191.12,195.49) --
	(189.32,193.66) --
	(187.51,191.78) --
	(185.71,189.84) --
	(183.90,187.83) --
	(182.10,185.74) --
	(180.29,183.57) --
	(178.49,181.31) --
	(176.68,178.96) --
	(174.88,176.50) --
	(173.08,173.93) --
	(171.27,171.26) --
	(169.47,168.48) --
	(167.66,165.58) --
	(165.86,162.58) --
	(164.05,159.46) --
	(162.25,156.24) --
	(160.44,152.92) --
	(158.64,148.96) --
	(156.83,142.05) --
	(155.03,133.61) --
	(153.23,126.17) --
	(151.42,119.78) --
	(149.62,114.26) --
	(147.81,109.45) --
	(146.01,105.24) --
	(144.20,101.52) --
	(142.40, 98.20) --
	(140.59, 95.21) --
	(138.79, 92.50) --
	(136.98, 90.01) --
	(135.18, 87.70) --
	(133.37, 85.54) --
	(131.57, 83.49) --
	(129.77, 81.53) --
	(127.96, 79.64) --
	(126.16, 77.79) --
	(124.35, 75.98) --
	(122.55, 74.19) --
	(120.74, 72.42) --
	(118.94, 70.67) --
	(117.13, 68.97) --
	(115.33, 67.37) --
	(113.52, 65.89) --
	(111.72, 64.55) --
	(109.92, 63.38) --
	(108.11, 62.36) --
	(106.31, 61.45) --
	(104.50, 60.62) --
	(102.70, 59.85) --
	(100.89, 59.10) --
	( 99.09, 58.35) --
	( 97.28, 57.60) --
	( 95.48, 56.82) --
	( 93.67, 56.00) --
	( 91.87, 55.16) --
	( 90.07, 54.32) --
	( 88.26, 53.47) --
	( 86.46, 52.64) --
	( 84.65, 51.82) --
	( 82.85, 51.04) --
	( 81.04, 50.30) --
	( 79.24, 49.59) --
	( 77.43, 48.93) --
	( 75.63, 48.32) --
	( 73.82, 47.76) --
	( 72.02, 47.24) --
	( 70.21, 46.76) --
	( 68.41, 46.34) --
	( 66.61, 45.95) --
	( 64.80, 45.60) --
	( 63.00, 45.29) --
	( 61.19, 45.02) --
	( 59.39, 44.78) --
	( 57.58, 44.56) --
	( 55.78, 44.38) --
	( 53.97, 44.21) --
	( 52.17, 44.07) --
	( 50.36, 43.94) --
	( 48.56, 43.83) --
	( 46.76, 43.74) --
	( 44.95, 43.66) --
	( 43.15, 43.59);
\definecolor{drawColor}{RGB}{0,0,255}

\path[draw=drawColor,line width= 1.2pt,line join=round,line cap=round] (178.64, 41.08) --
	(178.64, 45.21);
\end{scope}
\begin{scope}
\path[clip] (  0.00,  0.00) rectangle (252.94,252.94);
\definecolor{drawColor}{RGB}{0,0,0}

\path[draw=drawColor,line width= 0.4pt,line join=round,line cap=round] ( 43.15, 43.15) -- (221.80, 43.15);

\path[draw=drawColor,line width= 0.4pt,line join=round,line cap=round] ( 43.15, 43.15) -- ( 43.15, 37.15);

\path[draw=drawColor,line width= 0.4pt,line join=round,line cap=round] ( 78.88, 43.15) -- ( 78.88, 37.15);

\path[draw=drawColor,line width= 0.4pt,line join=round,line cap=round] (114.61, 43.15) -- (114.61, 37.15);

\path[draw=drawColor,line width= 0.4pt,line join=round,line cap=round] (150.34, 43.15) -- (150.34, 37.15);

\path[draw=drawColor,line width= 0.4pt,line join=round,line cap=round] (186.07, 43.15) -- (186.07, 37.15);

\path[draw=drawColor,line width= 0.4pt,line join=round,line cap=round] (221.80, 43.15) -- (221.80, 37.15);

\node[text=drawColor,anchor=base,inner sep=0pt, outer sep=0pt, scale=  1.00] at ( 43.15, 27.55) {0.0};

\node[text=drawColor,anchor=base,inner sep=0pt, outer sep=0pt, scale=  1.00] at ( 78.88, 27.55) {0.2};

\node[text=drawColor,anchor=base,inner sep=0pt, outer sep=0pt, scale=  1.00] at (114.61, 27.55) {0.4};

\node[text=drawColor,anchor=base,inner sep=0pt, outer sep=0pt, scale=  1.00] at (150.34, 27.55) {0.6};

\node[text=drawColor,anchor=base,inner sep=0pt, outer sep=0pt, scale=  1.00] at (186.07, 27.55) {0.8};

\node[text=drawColor,anchor=base,inner sep=0pt, outer sep=0pt, scale=  1.00] at (221.80, 27.55) {1.0};

\path[draw=drawColor,line width= 0.4pt,line join=round,line cap=round] ( 43.15, 43.15) -- ( 43.15,181.00);

\path[draw=drawColor,line width= 0.4pt,line join=round,line cap=round] ( 43.15, 43.15) -- ( 37.15, 43.15);

\path[draw=drawColor,line width= 0.4pt,line join=round,line cap=round] ( 43.15,112.07) -- ( 37.15,112.07);

\path[draw=drawColor,line width= 0.4pt,line join=round,line cap=round] ( 43.15,181.00) -- ( 37.15,181.00);

\node[text=drawColor,anchor=base east,inner sep=0pt, outer sep=0pt, scale=  1.00] at ( 31.15, 39.70) {0};

\node[text=drawColor,anchor=base east,inner sep=0pt, outer sep=0pt, scale=  1.00] at ( 31.15,108.63) {1};

\node[text=drawColor,anchor=base east,inner sep=0pt, outer sep=0pt, scale=  1.00] at ( 31.15,177.56) {2};

\path[draw=drawColor,line width= 0.4pt,line join=round,line cap=round] ( 43.15, 43.15) -- ( 43.15,228.95);

\node[text=drawColor,anchor=base,inner sep=0pt, outer sep=0pt, scale=  1.00] at (132.47, 14.40) {$H$};
\end{scope}
\end{tikzpicture}
\caption{With 3 land covers}
\end{subfigure}
\begin{subfigure}{0.45\columnwidth}
  \centering
\begin{tikzpicture}[x=1pt,y=1pt]
\definecolor{fillColor}{RGB}{255,255,255}
\path[use as bounding box,fill=fillColor,fill opacity=0.00] (25,10) rectangle (228.94,228.94);
\begin{scope}
\path[clip] (  0.00,  0.00) rectangle (252.94,252.94);
\definecolor{drawColor}{RGB}{0,0,0}

\node[text=drawColor,rotate= 90.00,anchor=base,inner sep=0pt, outer sep=0pt, scale=  1.00] at ( -2.40,132.47) {Density};
\end{scope}
\begin{scope}
\path[clip] ( 36.00, 36.00) rectangle (228.94,228.94);
\definecolor{drawColor}{gray}{0.40}

\path[draw=drawColor,line width= 0.4pt,line join=round,line cap=round] ( 43.15, 43.15) rectangle ( 52.08, 43.15);

\path[draw=drawColor,line width= 0.4pt,line join=round,line cap=round] ( 52.08, 43.15) rectangle ( 61.01, 43.21);

\path[draw=drawColor,line width= 0.4pt,line join=round,line cap=round] ( 61.01, 43.15) rectangle ( 69.94, 43.40);

\path[draw=drawColor,line width= 0.4pt,line join=round,line cap=round] ( 69.94, 43.15) rectangle ( 78.88, 43.83);

\path[draw=drawColor,line width= 0.4pt,line join=round,line cap=round] ( 78.88, 43.15) rectangle ( 87.81, 44.59);

\path[draw=drawColor,line width= 0.4pt,line join=round,line cap=round] ( 87.81, 43.15) rectangle ( 96.74, 45.77);

\path[draw=drawColor,line width= 0.4pt,line join=round,line cap=round] ( 96.74, 43.15) rectangle (105.67, 47.66);

\path[draw=drawColor,line width= 0.4pt,line join=round,line cap=round] (105.67, 43.15) rectangle (114.61, 50.71);

\path[draw=drawColor,line width= 0.4pt,line join=round,line cap=round] (114.61, 43.15) rectangle (123.54, 54.97);

\path[draw=drawColor,line width= 0.4pt,line join=round,line cap=round] (123.54, 43.15) rectangle (132.47, 61.48);

\path[draw=drawColor,line width= 0.4pt,line join=round,line cap=round] (132.47, 43.15) rectangle (141.41, 72.39);

\path[draw=drawColor,line width= 0.4pt,line join=round,line cap=round] (141.41, 43.15) rectangle (150.34, 88.16);

\path[draw=drawColor,line width= 0.4pt,line join=round,line cap=round] (150.34, 43.15) rectangle (159.27,109.00);

\path[draw=drawColor,line width= 0.4pt,line join=round,line cap=round] (159.27, 43.15) rectangle (168.20,133.17);

\path[draw=drawColor,line width= 0.4pt,line join=round,line cap=round] (168.20, 43.15) rectangle (177.14,164.47);

\path[draw=drawColor,line width= 0.4pt,line join=round,line cap=round] (177.14, 43.15) rectangle (186.07,201.03);

\path[draw=drawColor,line width= 0.4pt,line join=round,line cap=round] (186.07, 43.15) rectangle (195.00,221.80);

\path[draw=drawColor,line width= 0.4pt,line join=round,line cap=round] (195.00, 43.15) rectangle (203.93,214.93);

\path[draw=drawColor,line width= 0.4pt,line join=round,line cap=round] (203.93, 43.15) rectangle (212.87,191.26);

\path[draw=drawColor,line width= 0.4pt,line join=round,line cap=round] (212.87, 43.15) rectangle (221.80,131.20);
\definecolor{drawColor}{RGB}{255,0,0}

\path[draw=drawColor,line width= 0.4pt,line join=round,line cap=round] (221.80, 71.90) --
	(219.99, 98.62) --
	(218.19,131.62) --
	(216.39,148.20) --
	(214.58,158.46) --
	(212.78,169.84) --
	(210.97,180.16) --
	(209.17,189.19) --
	(207.36,196.83) --
	(205.56,203.08) --
	(203.75,208.02) --
	(201.95,211.79) --
	(200.14,214.61) --
	(198.34,216.69) --
	(196.53,218.32) --
	(194.73,219.78) --
	(192.93,221.35) --
	(191.12,223.30) --
	(189.32,223.90) --
	(187.51,221.65) --
	(185.71,217.09) --
	(183.90,210.83) --
	(182.10,203.48) --
	(180.29,195.58) --
	(178.49,187.64) --
	(176.68,179.98) --
	(174.88,172.64) --
	(173.08,165.59) --
	(171.27,158.83) --
	(169.47,152.34) --
	(167.66,146.10) --
	(165.86,140.11) --
	(164.05,134.36) --
	(162.25,128.84) --
	(160.44,123.55) --
	(158.64,118.47) --
	(156.83,113.60) --
	(155.03,108.94) --
	(153.23,104.48) --
	(151.42,100.23) --
	(149.62, 96.16) --
	(147.81, 92.29) --
	(146.01, 88.61) --
	(144.20, 85.10) --
	(142.40, 81.69) --
	(140.59, 78.32) --
	(138.79, 75.01) --
	(136.98, 71.87) --
	(135.18, 69.01) --
	(133.37, 66.51) --
	(131.57, 64.40) --
	(129.77, 62.61) --
	(127.96, 61.06) --
	(126.16, 59.70) --
	(124.35, 58.47) --
	(122.55, 57.33) --
	(120.74, 56.26) --
	(118.94, 55.24) --
	(117.13, 54.25) --
	(115.33, 53.30) --
	(113.52, 52.40) --
	(111.72, 51.55) --
	(109.92, 50.75) --
	(108.11, 50.00) --
	(106.31, 49.30) --
	(104.50, 48.66) --
	(102.70, 48.07) --
	(100.89, 47.53) --
	( 99.09, 47.03) --
	( 97.28, 46.59) --
	( 95.48, 46.18) --
	( 93.67, 45.82) --
	( 91.87, 45.49) --
	( 90.07, 45.20) --
	( 88.26, 44.94) --
	( 86.46, 44.71) --
	( 84.65, 44.51) --
	( 82.85, 44.33) --
	( 81.04, 44.17) --
	( 79.24, 44.03) --
	( 77.43, 43.91) --
	( 75.63, 43.81) --
	( 73.82, 43.71) --
	( 72.02, 43.63) --
	( 70.21, 43.57) --
	( 68.41, 43.51) --
	( 66.61, 43.45) --
	( 64.80, 43.41) --
	( 63.00, 43.37) --
	( 61.19, 43.34) --
	( 59.39, 43.31) --
	( 57.58, 43.29) --
	( 55.78, 43.27) --
	( 53.97, 43.25) --
	( 52.17, 43.23) --
	( 50.36, 43.22) --
	( 48.56, 43.21) --
	( 46.76, 43.20) --
	( 44.95, 43.19) --
	( 43.15, 43.19);
\definecolor{drawColor}{RGB}{0,0,255}

\path[draw=drawColor,line width= 1.2pt,line join=round,line cap=round] (182.76, 41.43) --
	(182.76, 44.86);
\end{scope}
\begin{scope}
\path[clip] (  0.00,  0.00) rectangle (252.94,252.94);
\definecolor{drawColor}{RGB}{0,0,0}

\path[draw=drawColor,line width= 0.4pt,line join=round,line cap=round] ( 43.15, 43.15) -- (221.80, 43.15);

\path[draw=drawColor,line width= 0.4pt,line join=round,line cap=round] ( 43.15, 43.15) -- ( 43.15, 37.15);

\path[draw=drawColor,line width= 0.4pt,line join=round,line cap=round] ( 78.88, 43.15) -- ( 78.88, 37.15);

\path[draw=drawColor,line width= 0.4pt,line join=round,line cap=round] (114.61, 43.15) -- (114.61, 37.15);

\path[draw=drawColor,line width= 0.4pt,line join=round,line cap=round] (150.34, 43.15) -- (150.34, 37.15);

\path[draw=drawColor,line width= 0.4pt,line join=round,line cap=round] (186.07, 43.15) -- (186.07, 37.15);

\path[draw=drawColor,line width= 0.4pt,line join=round,line cap=round] (221.80, 43.15) -- (221.80, 37.15);

\node[text=drawColor,anchor=base,inner sep=0pt, outer sep=0pt, scale=  1.00] at ( 43.15, 27.55) {0.0};

\node[text=drawColor,anchor=base,inner sep=0pt, outer sep=0pt, scale=  1.00] at ( 78.88, 27.55) {0.2};

\node[text=drawColor,anchor=base,inner sep=0pt, outer sep=0pt, scale=  1.00] at (114.61, 27.55) {0.4};

\node[text=drawColor,anchor=base,inner sep=0pt, outer sep=0pt, scale=  1.00] at (150.34, 27.55) {0.6};

\node[text=drawColor,anchor=base,inner sep=0pt, outer sep=0pt, scale=  1.00] at (186.07, 27.55) {0.8};

\node[text=drawColor,anchor=base,inner sep=0pt, outer sep=0pt, scale=  1.00] at (221.80, 27.55) {1.0};

\path[draw=drawColor,line width= 0.4pt,line join=round,line cap=round] ( 43.15, 43.15) -- ( 43.15,214.63);

\path[draw=drawColor,line width= 0.4pt,line join=round,line cap=round] ( 43.15, 43.15) -- ( 37.15, 43.15);

\path[draw=drawColor,line width= 0.4pt,line join=round,line cap=round] ( 43.15,100.31) -- ( 37.15,100.31);

\path[draw=drawColor,line width= 0.4pt,line join=round,line cap=round] ( 43.15,157.47) -- ( 37.15,157.47);

\path[draw=drawColor,line width= 0.4pt,line join=round,line cap=round] ( 43.15,214.63) -- ( 37.15,214.63);

\node[text=drawColor,anchor=base east,inner sep=0pt, outer sep=0pt, scale=  1.00] at ( 31.15, 39.70) {0};

\node[text=drawColor,anchor=base east,inner sep=0pt, outer sep=0pt, scale=  1.00] at ( 31.15, 96.86) {1};

\node[text=drawColor,anchor=base east,inner sep=0pt, outer sep=0pt, scale=  1.00] at ( 31.15,154.03) {2};

\node[text=drawColor,anchor=base east,inner sep=0pt, outer sep=0pt, scale=  1.00] at ( 31.15,211.19) {3};

\node[text=drawColor,anchor=base,inner sep=0pt, outer sep=0pt, scale=  1.00] at (132.47, 14.40) {$H$};
\end{scope}
\end{tikzpicture}
\caption{With 4 land covers}
\end{subfigure}
\caption{Histogram and density approximation for the random Shannon index $H$
  for 3 and 4 land covers. The blue line corresponds to the mean of the sample data. 
  $H$ has been calculated from a simulated uniform sample of $p$ of size $10^6$.}
  \label{fig_H_den}
\end{figure}
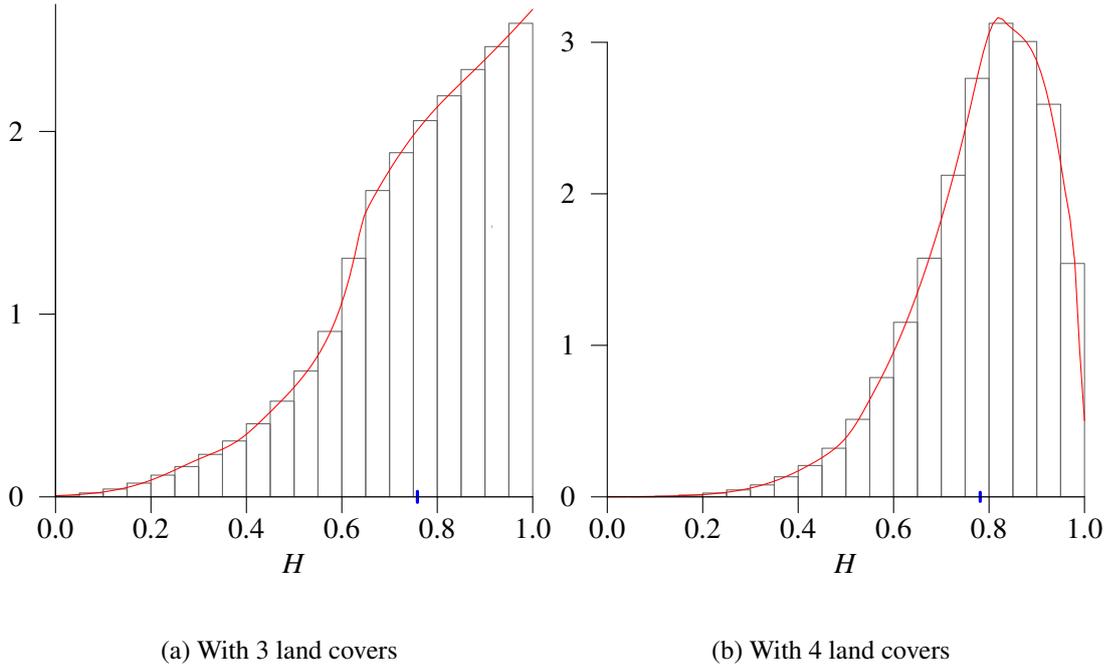

It is not possible to find analytically the probability 
distribution of $H$ from the law of
$p$. However it is trivial to generate random samples
of $p$ according to the uniform distribution on the simplex
and draw a histogram of values of $H$ using (\ref{eq:AHomega}). 
In Figure \ref{fig_H_den}, we show those histograms for 3 and 4 land covers,
obtained with a sample size of one million.
We have also added to the figure an estimation of the density function of $H$
and the position of the sample mean.

The density estimation has been carried out using the logsplines
method implemented in the R package \verb|logspline| 
\cite{kooperberg1992logspline}.
The usual kernel methods to estimate densities are not 
suitable here because $H$ is a bounded random variable.
 The base uniform sample on the simplex has been generated using the algorithm explained in \cite{rubinstein1998modern}:
 If $Y_1,\dots,Y_{n+1}$ are independent unit-exponential random variables, and 
 \begin{equation}
   E_i=\frac{Y_i}{\sum_{j=1}^{n+1}Y_j}
   \ ,
 \end{equation}
 then the random vector $(E_1,\dots,E_{n+1})$ is uniformly distributed on $\Delta$.


In fact, one does not need to estimate the theoretical mean of the distribution 
of $H$ by simulation, since it can be computed exactly. Indeed, 

the integral of $x\log_{n+1} x$ against the density (\ref{eq:p_marginal}), yields
\begin{equation*}
\frac{-1}{\ln (n+1)}\Big[\frac{\Psi(n+1)+\gamma-1}{n+1}+\frac{1}{(n+1)^2}\Big]
\ ,
\end{equation*}
where $\Psi$ is the digamma function, and $\gamma$ is the Euler-Mascheroni constant.
Therefore, the expectation of the Shannon index $H$ of (\ref{eq:AHomega}), under the hypothesis
of uniform distribution of the proportions $p_i$ in the simplex, 
is given by
\begin{equation*}
\E[H]=
\frac{1}{\ln (n+1)}\Big[\Psi(n+1)+\gamma-1+\frac{1}{n+1}\Big]
\ .
\end{equation*}
This expectation tends to 1 as $n\to\infty$, as it is easily seen from
the inequalities $\ln n\le \Psi(n+1)\le\ln (n+1)$.

\subsection{The distribution of $A$}

For $A$ it is possible, on the contrary, to deduce 
an analytical formula for its probability distribution,
because it is a simple linear function of the proportions
$p$. 

Without loss of generality, we can assume that the weights
$w=(w_1,\hdots, w_{n+1})$ are sorted and different: $0<w_1<\cdots<w_{n+1}$.
We can write
\begin{align*}
A&=\sum_{i=1}^{n}w_ip_i+w_{n+1}\Big(1-\sum_{i=1}^np_i\Big)\\
&=w_{n+1}-\sum_{i=1}^ns_ip_i,
\end{align*}
where $s_i:=w_{n+1}-w_i$, 
and clearly $0<s_n<s_{n-1}<\cdots<s_1<w_{n+1}$.

To obtain the distribution of $A$ when  
$p$ is uniform on $\Delta'$, let us compute first the probability
density of $\sum_{i=1}^ns_ip_i=w_{n+1}-A$.
We use a change of variable by means of the bijective linear transformation
  $T\colon \Delta'\longrightarrow B\subset\RR^n$
given by 
\begin{equation*}
\begin{cases}
  v_1=\sum_{i=1}^n s_i p_i & \\
  v_j=s_j p_j\ ,\quad j=2,\dots,n &
\end{cases}
\end{equation*}
where
\begin{equation*}
B=\Big\{v\in \RR^n:\ \Sum_{i=1}^n\frac{v_i}{s_i}-\Sum_{i=2}^n\frac{v_i}{s_1}\leq 1,\ \Sum_{i=2}^nv_i\leq v_1,\ \text{and } v_i\geq 0\Big\}\ .
\end{equation*}

The inverse mapping
  $T^{-1}\colon B\longrightarrow \Delta'$
is defined by 
\begin{equation*}
\begin{cases}
  p_1=\frac{1}{s_1}\big(v_1-\sum_{i=2}^n v_i\big) &  \\
  p_j=s_j v_j\ ,\quad j=2,\dots,n &
\end{cases}
\end{equation*}
with Jacobian determinant equal to $\prod_{i=1}^n \frac{1}{s_i}$.
Therefore, the density of the vector $ v=(v_1,\dots,v_n)$ is given by 
\begin{equation}\label{eq:dens_v}
f(v_1,\ldots,v_n)=\begin{cases}
n!\prod_{i=1}^n\frac{1}{s_i}& \text{ if $v\in B$} \\
0&\text{otherwise}\ .  
\end{cases}
\end{equation}

To obtain the density of $v_1$, we integrate (\ref{eq:dens_v}) with respect to $v_2,\dots,v_n$.
For fixed $v_{1},\dots,v_{k-1}$, the variable $v_k$ ranges from $0$ to $m_k$, with
\begin{equation*}
\textstyle
m_k=\min\Big\{v_1-\sum_{i=2}^{k-1}v_i ,\ 
\frac{s_1s_k}{s_1-s_k}\(1-\frac{v_1}{s_1}-\sum_{i=2}^{k-1}v_i\frac{s_1-s_i}{s_1s_i}\)\Big\}\ .
\end{equation*}
Hence,
\begin{equation*}
f(v_1)=\int_0^{m_2}\cdots\int_0^{m_n}
  n!\prod_{i=1}^n\frac{1}{s_i}\ dv_n\cdots dv_2\ ,
\end{equation*}
which can be exactly computed for given values of $s_1,\dots,s_n$.

\begin{figure}
  \begin{subfigure}{0.45\columnwidth}
    \centering
  \input{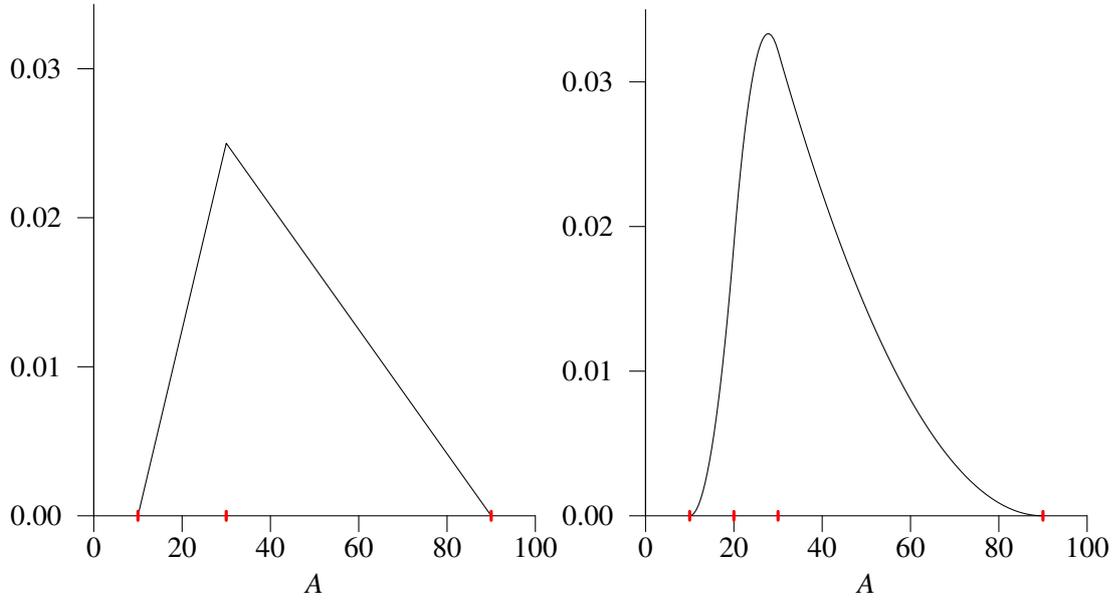}
    \caption{With 3 land covers and $w=(10, 30, 90)$}
  \end{subfigure}
  \begin{subfigure}{0.45\columnwidth}
    \centering
\begin{tikzpicture}[x=1pt,y=1pt]
\definecolor{fillColor}{RGB}{255,255,255}
\path[use as bounding box,fill=fillColor,fill opacity=0.00] (25,40) rectangle (227.75,261.56);
\begin{scope}
\path[clip] ( 49.20, 61.20) rectangle (227.75,261.56);
\definecolor{drawColor}{RGB}{0,0,0}

\path[draw=drawColor,line width= 0.4pt,line join=round,line cap=round] ( 72.34, 68.62) --
	( 73.17, 68.88) --
	( 74.00, 69.64) --
	( 74.82, 70.92) --
	( 75.65, 72.71) --
	( 76.48, 75.02) --
	( 77.30, 77.83) --
	( 78.13, 81.15) --
	( 78.96, 84.99) --
	( 79.78, 89.34) --
	( 80.61, 94.20) --
	( 81.44, 99.57) --
	( 82.26,105.45) --
	( 83.09,111.85) --
	( 83.92,118.75) --
	( 84.74,126.17) --
	( 85.57,134.10) --
	( 86.40,142.54) --
	( 87.22,151.49) --
	( 88.05,160.95) --
	( 88.88,170.93) --
	( 89.70,180.83) --
	( 90.53,190.08) --
	( 91.36,198.66) --
	( 92.18,206.59) --
	( 93.01,213.86) --
	( 93.84,220.48) --
	( 94.66,226.43) --
	( 95.49,231.73) --
	( 96.32,236.37) --
	( 97.14,240.35) --
	( 97.97,243.68) --
	( 98.80,246.35) --
	( 99.62,248.35) --
	(100.45,249.71) --
	(101.28,250.40) --
	(102.10,250.44) --
	(102.93,249.82) --
	(103.76,248.54) --
	(104.58,246.60) --
	(105.41,244.01) --
	(106.24,241.10) --
	(107.06,238.21) --
	(107.89,235.35) --
	(108.72,232.51) --
	(109.54,229.70) --
	(110.37,226.91) --
	(111.19,224.14) --
	(112.02,221.40) --
	(112.85,218.69) --
	(113.67,215.99) --
	(114.50,213.33) --
	(115.33,210.68) --
	(116.15,208.06) --
	(116.98,205.47) --
	(117.81,202.90) --
	(118.63,200.36) --
	(119.46,197.83) --
	(120.29,195.34) --
	(121.11,192.86) --
	(121.94,190.42) --
	(122.77,187.99) --
	(123.59,185.59) --
	(124.42,183.22) --
	(125.25,180.87) --
	(126.07,178.54) --
	(126.90,176.24) --
	(127.73,173.96) --
	(128.55,171.71) --
	(129.38,169.48) --
	(130.21,167.28) --
	(131.03,165.10) --
	(131.86,162.94) --
	(132.69,160.81) --
	(133.51,158.70) --
	(134.34,156.62) --
	(135.17,154.56) --
	(135.99,152.53) --
	(136.82,150.52) --
	(137.65,148.53) --
	(138.47,146.57) --
	(139.30,144.63) --
	(140.13,142.72) --
	(140.95,140.83) --
	(141.78,138.97) --
	(142.61,137.13) --
	(143.43,135.32) --
	(144.26,133.53) --
	(145.09,131.76) --
	(145.91,130.02) --
	(146.74,128.30) --
	(147.57,126.61) --
	(148.39,124.94) --
	(149.22,123.29) --
	(150.04,121.68) --
	(150.87,120.08) --
	(151.70,118.51) --
	(152.52,116.96) --
	(153.35,115.44) --
	(154.18,113.94) --
	(155.00,112.47) --
	(155.83,111.02) --
	(156.66,109.59) --
	(157.48,108.19) --
	(158.31,106.82) --
	(159.14,105.46) --
	(159.96,104.14) --
	(160.79,102.83) --
	(161.62,101.55) --
	(162.44,100.30) --
	(163.27, 99.07) --
	(164.10, 97.86) --
	(164.92, 96.68) --
	(165.75, 95.53) --
	(166.58, 94.39) --
	(167.40, 93.28) --
	(168.23, 92.20) --
	(169.06, 91.14) --
	(169.88, 90.11) --
	(170.71, 89.09) --
	(171.54, 88.11) --
	(172.36, 87.15) --
	(173.19, 86.21) --
	(174.02, 85.29) --
	(174.84, 84.41) --
	(175.67, 83.54) --
	(176.50, 82.70) --
	(177.32, 81.88) --
	(178.15, 81.09) --
	(178.98, 80.33) --
	(179.80, 79.58) --
	(180.63, 78.86) --
	(181.46, 78.17) --
	(182.28, 77.50) --
	(183.11, 76.85) --
	(183.94, 76.23) --
	(184.76, 75.64) --
	(185.59, 75.06) --
	(186.42, 74.52) --
	(187.24, 73.99) --
	(188.07, 73.49) --
	(188.89, 73.02) --
	(189.72, 72.57) --
	(190.55, 72.14) --
	(191.37, 71.74) --
	(192.20, 71.36) --
	(193.03, 71.01) --
	(193.85, 70.68) --
	(194.68, 70.37) --
	(195.51, 70.09) --
	(196.33, 69.84) --
	(197.16, 69.61) --
	(197.99, 69.40) --
	(198.81, 69.22) --
	(199.64, 69.06) --
	(200.47, 68.93) --
	(201.29, 68.82) --
	(202.12, 68.73) --
	(202.95, 68.67) --
	(203.77, 68.63) --
	(204.60, 68.62);
\end{scope}
\begin{scope}
\path[clip] (  0.00,  0.00) rectangle (252.94,310.76);
\definecolor{drawColor}{RGB}{0,0,0}

\path[draw=drawColor,line width= 0.4pt,line join=round,line cap=round] ( 55.81, 68.62) -- (221.13, 68.62);

\path[draw=drawColor,line width= 0.4pt,line join=round,line cap=round] ( 55.81, 68.62) -- ( 55.81, 62.62);

\path[draw=drawColor,line width= 0.4pt,line join=round,line cap=round] ( 88.88, 68.62) -- ( 88.88, 62.62);

\path[draw=drawColor,line width= 0.4pt,line join=round,line cap=round] (121.94, 68.62) -- (121.94, 62.62);

\path[draw=drawColor,line width= 0.4pt,line join=round,line cap=round] (155.00, 68.62) -- (155.00, 62.62);

\path[draw=drawColor,line width= 0.4pt,line join=round,line cap=round] (188.07, 68.62) -- (188.07, 62.62);

\path[draw=drawColor,line width= 0.4pt,line join=round,line cap=round] (221.13, 68.62) -- (221.13, 62.62);

\node[text=drawColor,anchor=base,inner sep=0pt, outer sep=0pt, scale=  1.00] at ( 55.81, 53.02) {0};

\node[text=drawColor,anchor=base,inner sep=0pt, outer sep=0pt, scale=  1.00] at ( 88.88, 53.02) {20};

\node[text=drawColor,anchor=base,inner sep=0pt, outer sep=0pt, scale=  1.00] at (121.94, 53.02) {40};

\node[text=drawColor,anchor=base,inner sep=0pt, outer sep=0pt, scale=  1.00] at (155.00, 53.02) {60};

\node[text=drawColor,anchor=base,inner sep=0pt, outer sep=0pt, scale=  1.00] at (188.07, 53.02) {80};

\node[text=drawColor,anchor=base,inner sep=0pt, outer sep=0pt, scale=  1.00] at (221.13, 53.02) {100};

\path[draw=drawColor,line width= 0.4pt,line join=round,line cap=round] ( 55.81, 68.62) -- ( 55.81,232.31);

\path[draw=drawColor,line width= 0.4pt,line join=round,line cap=round] ( 55.81, 68.62) -- ( 49.81, 68.62);

\path[draw=drawColor,line width= 0.4pt,line join=round,line cap=round] ( 55.81,123.19) -- ( 49.81,123.19);

\path[draw=drawColor,line width= 0.4pt,line join=round,line cap=round] ( 55.81,177.75) -- ( 49.81,177.75);

\path[draw=drawColor,line width= 0.4pt,line join=round,line cap=round] ( 55.81,232.31) -- ( 49.81,232.31);

\node[text=drawColor,anchor=base east,inner sep=0pt, outer sep=0pt, scale=  1.00] at ( 43.81, 65.18) {0.00};

\node[text=drawColor,anchor=base east,inner sep=0pt, outer sep=0pt, scale=  1.00] at ( 43.81,119.74) {0.01};

\node[text=drawColor,anchor=base east,inner sep=0pt, outer sep=0pt, scale=  1.00] at ( 43.81,174.31) {0.02};

\node[text=drawColor,anchor=base east,inner sep=0pt, outer sep=0pt, scale=  1.00] at ( 43.81,228.87) {0.03};

\path[draw=drawColor,line width= 0.4pt,line join=round,line cap=round] ( 55.81, 68.62) -- ( 55.81,259.60);

\node[text=drawColor,anchor=base,inner sep=0pt, outer sep=0pt, scale=  1.00] at (138.47, 39.60) {$A$};
\end{scope}
\begin{scope}
\path[clip] ( 49.20, 61.20) rectangle (227.75,261.56);
\definecolor{drawColor}{RGB}{255,0,0}

\path[draw=drawColor,line width= 1.2pt,line join=round,line cap=round] ( 72.34, 66.98) --
	( 72.34, 70.26);

\path[draw=drawColor,line width= 1.2pt,line join=round,line cap=round] ( 88.88, 66.98) --
	( 88.88, 70.26);

\path[draw=drawColor,line width= 1.2pt,line join=round,line cap=round] (105.41, 66.98) --
	(105.41, 70.26);

\path[draw=drawColor,line width= 1.2pt,line join=round,line cap=round] (204.60, 66.98) --
	(204.60, 70.26);
\end{scope}
\end{tikzpicture}
    \caption{With 4 land covers and $w=(10, 20, 30, 90)$}
  \end{subfigure}
  \caption{Density of the appropriation $A$ for 3 and 4 land covers, and for a particular vector
   of weights $w$, with values indicated by the red marks.}
  \label{fig_hanpp}
\end{figure}

Finally, the density function of A is simply
\begin{equation*}
f_{A}(a)=
\begin{cases}
f(w_{n+1}-a)& \text{ if $a\in[w_1, w_{n+1}]$} \\
0& \text{ otherwise .}
\end{cases}
\end{equation*}
The graph of this function of $a$ is depicted in 
Figure \ref{fig_hanpp} for three and four land covers and some given values of $w$. 

%
%

The expected value of $A$ is easily computed using (\ref{eq:p_marginal}) directly,
or reasoned by symmetry:
\begin{equation*}
\E[A]=\frac{1}{n+1} \Sum_{i=1}^{n+1} w_i\ .
\end{equation*}

\subsection{Expected value of $H$ for a given appropriation}
We show in this subsection that a closed formula can be derived
for the expected value of the Shannon index $H$ conditioned to a given level
of appropriation $A$. Specifically, we want to compute the function
\begin{equation}\label{eq:HcondA}
  a\mapsto \E[ H \mid A=a]
\end{equation}

Since both $H$ and $A$ are functions of the vector of land covers 
$p=(p_1,\dots,p_n)\in\Delta'$, the conditional expectation can 
be computed by means of the conditional law of $p$ 
given $A(p)=a$.

\begin{lema-n} 
  Let $X=(X_1,\dots,X_n)$ be a random vector
  following a continuous uniform distribution with support on
  a Borel set $\Gamma\subset\RR^n$ and let 
  $Y:=\alpha_0+\alpha_1X_1+\cdots+\alpha_nX_n$, for some constants $\alpha_i\in\RR$.
  
  Then, the conditional distribution of $X$ given $\{Y=a\}$
  is uniform in $\RR^{n-1}$ with support on the intersection 
  $I_{a}:=\Gamma\cap\{\alpha_0+\alpha_1x_1+\cdots+\alpha_nx_n=a\}$,
  for almost all $a$ with respect to the law of $Y$.
\end{lema-n}

The fact stated in the lemma looks intuitive and it is indeed straightforward to prove.
Notice, however, that the fact that $\{\alpha_0+\alpha_1x_1+\cdots+\alpha_nx_n=a\}$
is a bundle of parallel lines is crucial, and that the result does not
say anything about a particular value $a$, but should be understood
with respect to the set of values $a$ as a whole.

We apply the lemma to $X=(p_1,\dots,p_n)$, $\Gamma=\Delta'$, and 
$Y=A=w_{n+1}-\Sum_{i=1}^n (w_{n+1}-w_i)p_i$.

The intersection of the simplex $\Delta'$ with the line $\{A=a\}$
is given by 
\begin{equation*}
  I_a=\{(p_1,\dots,p_{n-1})\in\RR^{n-1}:\ m_{k,a}\le p_k\le M_{k,a},\ \forall k\}\ ,
\end{equation*}
where 
\begin{align*}
 m_{k,a}:=&\max\Bigg\{0, \tfrac{w_{k+1}-a-\Sum_{i=1}^{k-1}(w_{k+1}-w_i)p_i}{w_{k+1}-w_k}\Bigg\}\ , \\
M_{k,a}:=&\tfrac{w_{n+1}-a-\Sum_{i=1}^{k-1}(w_{n+1}-w_i)p_i}{w_{n+1}-w_k}\ .
\end{align*}

Taking into account that, on $I_{a}$, we can write $p_n$ and $p_{n-1}$ as a function of the 
other coordinates, namely,
\begin{equation*}
  p_n=\tfrac{w_{n+1}-a-\sum_{i=1}^{n-1}(w_{n+1}-w_i)p_i}{w_{n+1}-w_n}
\end{equation*}
and
\begin{equation*}
  p_{n+1}=1-\sum_{i=1}^{n}p_i=\tfrac{a-w_n+\sum_{i=1}^{n-1}p_i(w_n-w_i)}{w_{n+1}-w_n}
  \ ,
\end{equation*}
we have that the conditional expectation (\ref{eq:HcondA}) is in fact
a function of $n-1$ coordinates of $p$, and can be expressed as  
\begin{equation}\label{eq:EHAa}
  \E[ H \mid A=a] =\int_{I_a} \textstyle{C_a^{-1} \Big[-\sum_{i=1}^{n+1}p_i\log_{n+1}p_i}\Big]\,dp\ ,
\end{equation} 
where
\begin{equation*}
  C_a:=\int_{m_{1,a}}^{M_{1,a}}\cdots\int_{m_{n-1,a}}^{M_{n-1,a}}\ dp_{n-1}\cdots dp_1
\end{equation*}
is the volume of $I_a$.

\begin{figure}[h!]
  \begin{subfigure}{0.45\columnwidth}
    \centering
  \begin{overpic}[scale=0.7,unit=1mm]{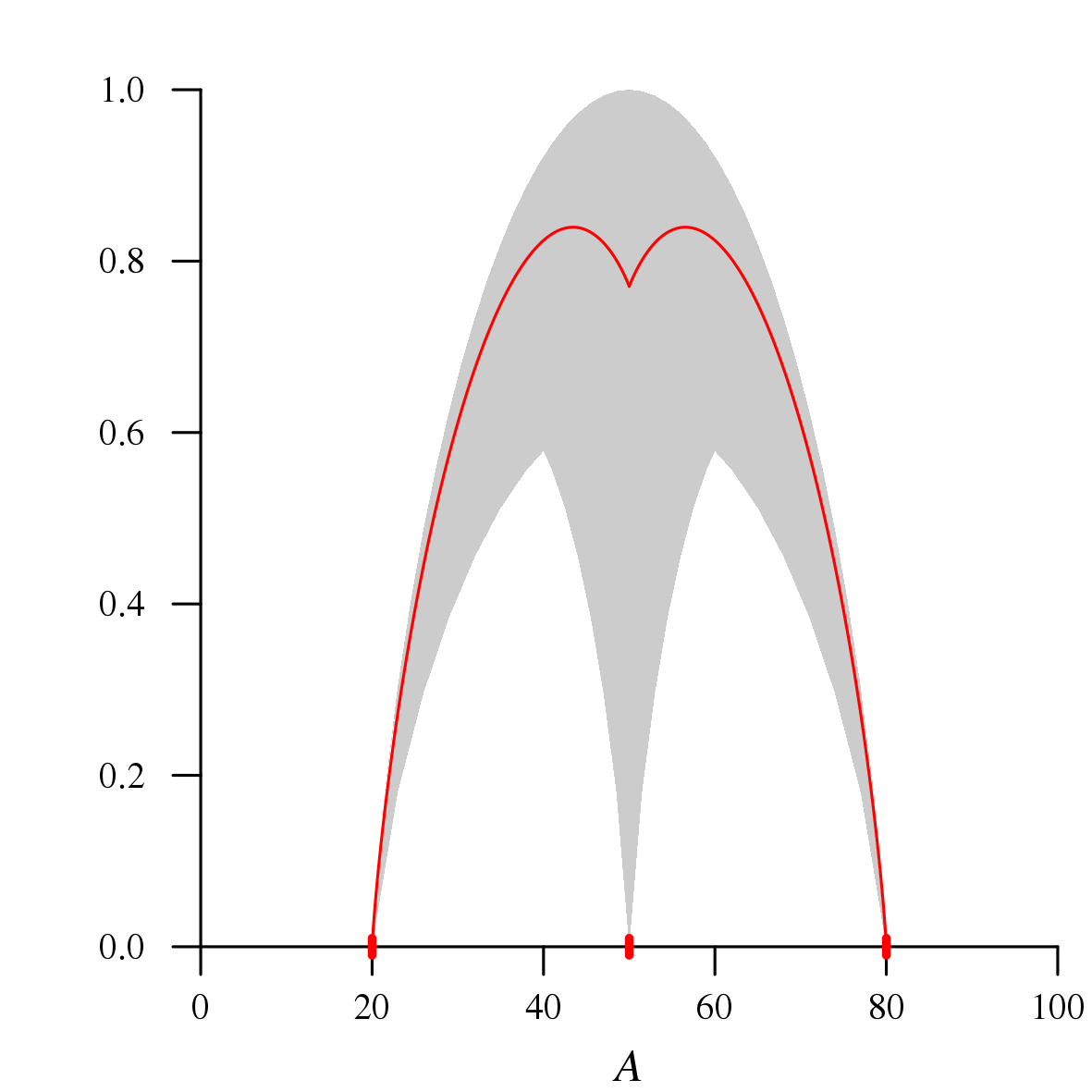}
   \put(-1,-1){\begin{turn}{90}
   {\parbox{1\linewidth}{
      \begin{equation*}
      \E[ H \mid A]
      \end{equation*}}}
      \end{turn}}
 \end{overpic}
   \caption{With weights $w=(20,50,80)$}
  \end{subfigure}
  \begin{subfigure}{0.45\columnwidth}
    \centering
  \begin{overpic}[scale=0.7,unit=1mm]{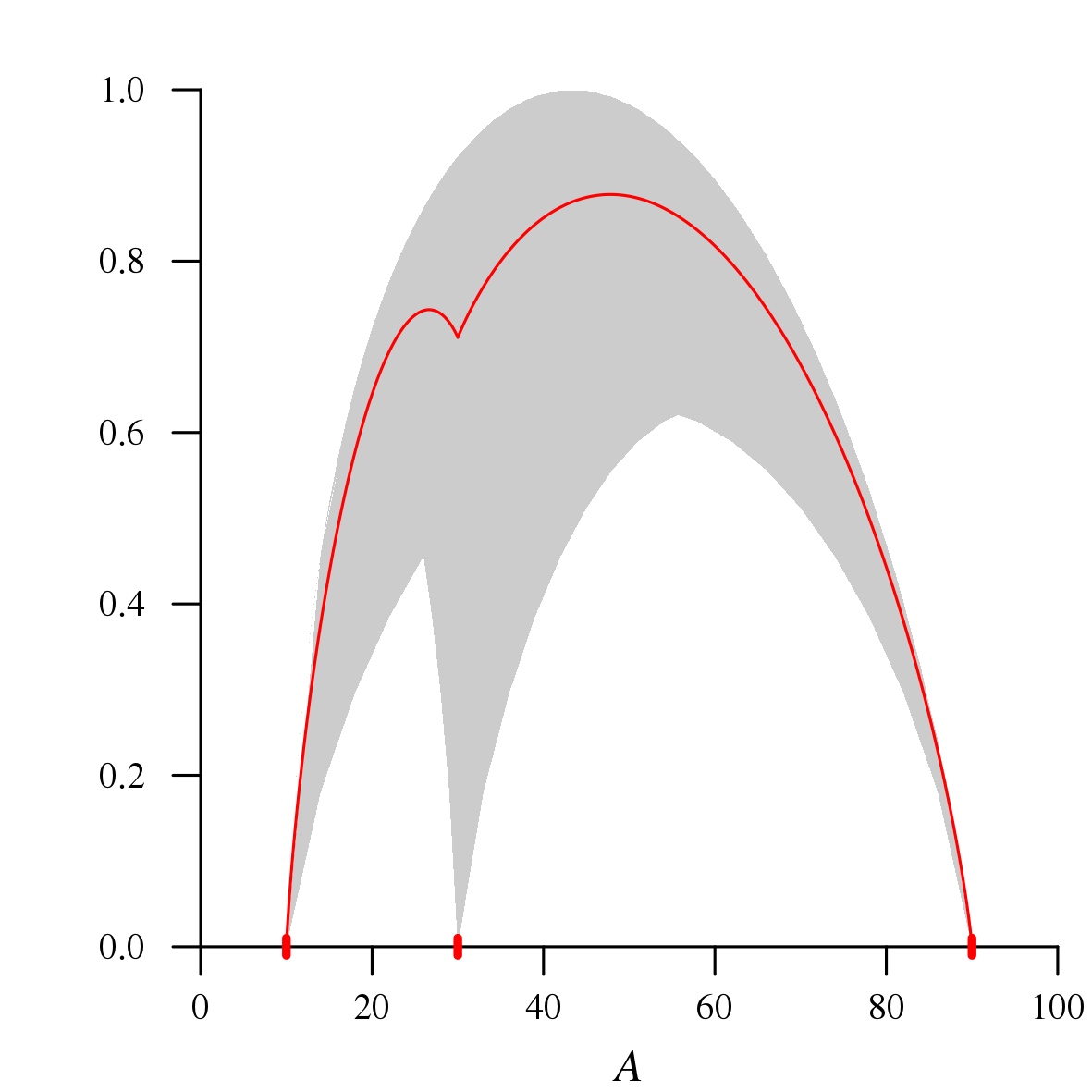}
   \put(-1,-1){\begin{turn}{90}
   {\parbox{1\linewidth}{
      \begin{equation*}
      \E[ H \mid A]
      \end{equation*}}}
      \end{turn}}
 \end{overpic}
    \caption{With weights $w=(10, 30, 90)$}
  \end{subfigure}
  \caption{The red curve is the expected value of the Shannon index $H$ as a function
  of the human appropriation $A$, for $n+1=3$ covers.
  The shaded area corresponds to the set of possible pairs of values $(A,H)$, and has been
  drawn by simulating one million points from its joint probability distribution.}
  \label{fig_espH_3}
\end{figure}

\begin{figure}[h!]
  \begin{subfigure}{0.45\columnwidth}
    \centering
  \begin{overpic}[scale=0.7,unit=1mm]{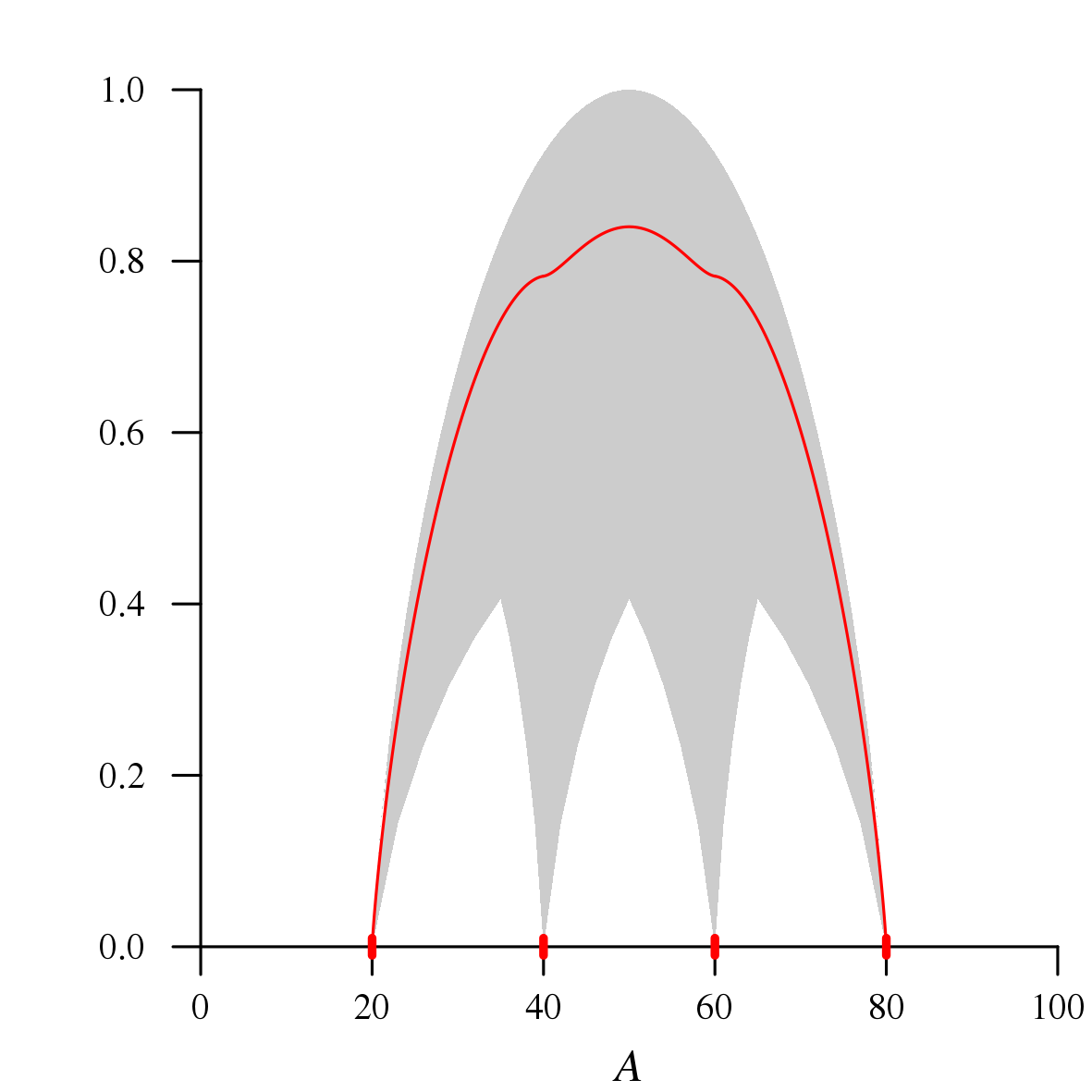}
   \put(-1,-1){\begin{turn}{90}
   {\parbox{1\linewidth}{
      \begin{equation*}
      \E[ H \mid A]
      \end{equation*}}}
      \end{turn}}
 \end{overpic}
    \caption{With weights $w=(20,40,60,80)$}
  \end{subfigure}
  \begin{subfigure}{0.45\columnwidth}
    \centering
  \begin{overpic}[scale=0.7,unit=1mm]{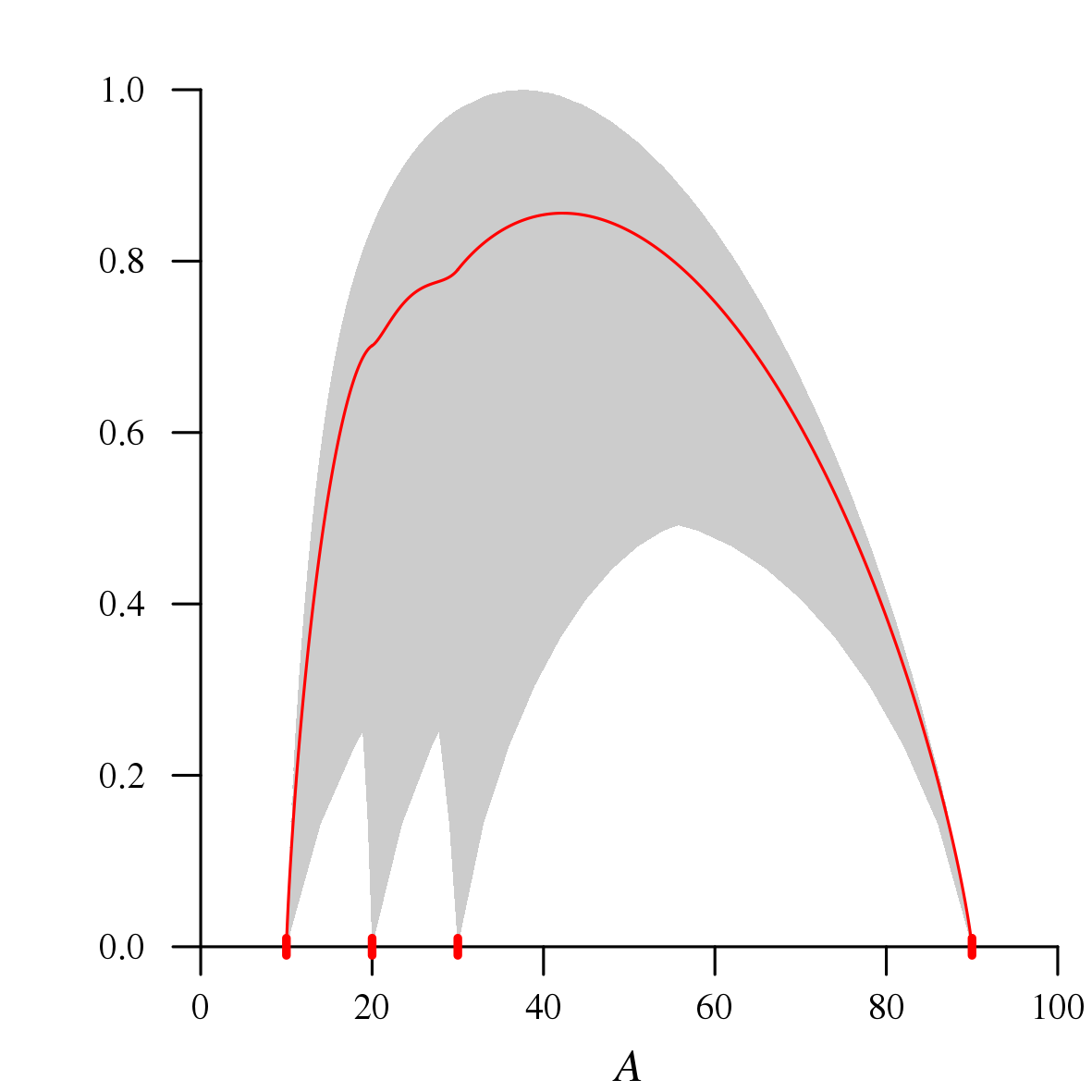}
   \put(-1,-1){\begin{turn}{90}
   {\parbox{1\linewidth}{
      \begin{equation*}
      \E[ H \mid A]
      \end{equation*}}}
      \end{turn}}
 \end{overpic}
    \caption{With weights $w=(10, 20, 30, 90)$}
  \end{subfigure}
  \caption{The analogues of Fig.~\ref{fig_espH_3} for $n+1=4$ covers, and the indicated
  weights. Note that the set of possible points and the conditional expectation curves
  are symmetrical if the weights are equidistant.}
  \label{fig_espH_4}
\end{figure}

The integral (\ref{eq:EHAa}) can be computed exactly as a piecewise function that depends on the value of $a$.
The result is given in Figures \ref{fig_espH_3} i \ref{fig_espH_4}, for $n+1=3$ and $n+1=4$ and two
sets of weights $w$. 
In all cases and dimensions the function (\ref{eq:HcondA}) is continuous, piecewise
concave, and non-smooth at the points $w_i$.

\section{Shannon index and appropriation with real data}\label{sec:realdata}

For the sake of simplicity, in this section we change $n+1$ to $n$ and hereinafter the simplex will be
$$\Delta=\left\{(p_1,\hdots,p_{n})\ |\ p_i\geq0,\ p_1+\cdots+p_{n}=1\right\} \ .$$

Given a wide region, divided in small cells, the proportion of 
land covers in each cell will rarely be well represented by the
uniform distribution. Not only some land covers can take more 
surface than others in the region, but also not all covers will be
present in all cells.

To apply the method of the previous section with sample data, 
we need first to estimate from the data the probability distribution
of land covers for the target region. 
Then, a large
sample will be drawn from that distribution, and the conditional 
expectation $\E[H\mid A=a]$ will be estimated from that sample. 
The analytical exact computation is of course no longer possible,
since there is no a closed analytical expression for the distribution
of $A$,
unlike the uniform case. 
 However, 
 the estimated distribution of the proportions $p$ allows to simulate
 as many values of $H$ and $A$ as desired, and these in
 turn allow to approximate $\E[H\mid A=a]$. The quality of the result
 depends on the quality of the estimation of the distribution of $p$ and on the
 number of values simulated.
 
This programme has some difficulties, that will be addressed
  in different subsections below. First, we develop the estimation of a density on 
  a simplex by means of Dirichlet kernels. 
  This estimation has numerical difficulties, that we solve in the second 
  subsection. Next, we consider the global sampling strategy, taking into account
  the many points that lie in the facets of the simplex, which are themselves simplices of
  lower dimensions. Finally, we explain our procedure to choose the
  bandwidth parameter of the kernels, an important detail that will be postponed in the first subsection.
  
  An option to avoid the difficulties with de Dirichlet kernels is to 
  employ the log-ratios 
  $y_i=\log(p_i/p_{n+1})$, see \cite{aitchison1985},
  or symmetric and isometric 
  log-ratios, see \cite{Chacon2011702}, 
  and then use kernels with unbounded domain, but   
  these methods have serious drawbacks 
  with samples whose points can very well be on the boundary of the simplex, as is in our case.

\subsection{Kernel density estimation on the simplex}\label{subsec:distp}
The estimation of probability distributions from data can be done in two ways:
Either postulating a parametric family of distributions and 
estimating the parameters from the data, or by letting the data
directly shape the distribution. 
In the second case, a probability \emph{density function} is usually assumed to exist,
and we speak of 
\emph{non-parametric density estimation}.

We dismissed the first method due to the following reason: 
the only standard family of distributions with bounded support
is the Dirichlet family,
but we found that our data was far from being well represented
by any of its members.
Nevertheless we will use the Dirichlet family in a different way,
as kernels to apply the \emph{kernel density estimation} method.
For the reader convenience, we recall here the definition of the
Dirichlet family and the kernel method:

The density function of the Dirichlet distribution of dimension $n>1$ and positive parameters 
$\alpha=(\alpha_1,\ldots, \alpha_n)$ is
\begin{equation}\label{eq:fDirichlet}
f(x_1,\ldots, x_n)=\frac{1}{B(\alpha)}\prod_{j=1}^n x_j^{\alpha_j-1}\ ,
\end{equation}
supported by the simplex $\Delta$, where $B$ is the multivariate Beta function:
\begin{equation*}
B(\alpha)=\frac{\prod_{j=1}^n\Gamma(\alpha_j)}{\Gamma(\sum_{j=1}^n\alpha_j)}\ ,
\quad\text{and}\quad
\Gamma(t)=\int_0^\infty x^{t-1}e^{-x}dx\ .
\end{equation*}

The kernel method, in general, consists of estimating the true density function $f$ by
\begin{equation*}
\hat{f}(x)=\frac{1}{N}\Sum_{i=1}^NK(x, z_i, \Lambda)\ ,
\end{equation*}
where $K$ is the \emph{kernel function}, which is a probability density function in $x$ 
depending on the sample 
points $z_i,\ i=1,\dots, N$,  
and on an $n\times n$ 
symmetric and positive-definite matrix $\Lambda$, called the \emph{smoothing} or \emph{bandwidth} matrix. 
As a function of $x$, $K$ attains its maximum at $x=z_i$. 
Parameters outside the diagonal in $\Lambda$ define the degree of covariance between the kernel marginal 
laws, and the size of its eigenvalues are related to the kernel spread, that is, the greater the eigenvalues, the larger 
the spread in the corresponding eigenvector direction. In general, the kernel methods have good asymptotic properties.

In the absence of any relevant additional information, we will take $\Lambda$ as a diagonal matrix
with the same variance $\lambda$ in all coordinate directions, and in consequence the kernel
will be the Dirichlet density (\ref{eq:fDirichlet}) with 
\begin{equation*}
\alpha_j=1+\frac{z_{ij}}{\lambda}\ ,
\end{equation*}
where $z_{ij}$ is the $j$-th coordinate of $z_i$.

Using a kernel supported on the simplex $\Delta$ ensures that the estimation is also supported
on $\Delta$. 
The choice of the bandwidth parameter $\lambda$
is crucial for an accurate estimation of the density. We have spent a considerable effort 
to get it right, and this is the contents of Subsection \ref{subsec:lambdafit}.

According with the assumptions above, our estimated density of the proportions $p$ 
is given by
\begin{align}\label{eq:estdens}
 \hatf(x)=&\ \frac{1}{N}\sum_{i=1}^N
 \frac{\Gamma\(n+\frac{1}{\lambda}\)}{\Prod_{j=1}^n\Gamma\(1+\frac{\zij}{\lambda}\)}
 \Prod_{j=1}^nx_j^{\zij/\lambda}\ .
 \end{align}

As we will see, in the search of the optimal value of $\lambda$, we will need to evaluate
(\ref{eq:estdens}) with $\lambda$ in the order of $10^{-3}$. That means, the gamma functions 
in both numerator and denominator will have a very large argument, with a subsequent loss of precision.
 For that reason, in Subsection \ref{subsec:numapprox} we look for an approximation 
 of the gamma function to simplify the quotient before evaluating each part. 

All of the above can be applied under the assumption that
there exists a density on the simplex. In our case this is in fact not true,
because there are data points in the lower
dimensional facets of the simplex, corresponding to the 
cells on which not all land covers are present. We explain the solution in
Subsection \ref{subsec:metod}.

\subsection{Numerical approximation of the estimated density}
\label{subsec:numapprox}
To get an appropriate numerical approximation of the quotient of gammas in (\ref{eq:estdens}),
we use Weierstrass' formula
 \begin{equation*}
 \Gamma(t+1)=e^{-\gamma t}\Prod_{k=1}^{\infty}(1+t/k)^{-1}e^{t/k}\ ,
 \end{equation*}
 where $\gamma$ is the Euler-Mascheroni constant again.
 Denoting
 \begin{equation*}
C:=\prod_{j=1}^{n-1}(n-j+\tfrac{1}{\lambda})\ ,
 \end{equation*}
the quotient in (\ref{eq:estdens}) can be written
\begin{align*}
  &\frac{\Gamma(n+\frac{1}{\lambda})}{\Prod_{j=1}^n\Gamma(1+\frac{\zij}{\lambda})}=
  C
  \Prod_{j=1}^n\frac{\Gamma(1+\frac{1}{\lambda})^{1/n}}{\Gamma(1+\frac{\zij}{\lambda})}
  \\
  &=C\Prod_{j=1}^n\Big[e^{-\frac{\gamma}{\lambda}(\frac{1}{n}-\zij)}
  \Prod_{k=1}^\infty\frac{1+\frac{\zij}{k\lambda}}{(1+\frac{1}{k\lambda})^{\frac{1}{n}}}
  e^{\frac{1}{k\lambda}(\frac{1}{n}-\zij)}\Big]
  \\
  &=C\Prod_{j=1}^n\Big[e^{-\frac{\gamma}{\lambda}(\frac{1}{n}-\zij)}
  \exp\Big\{\Sum_{k=1}^\infty\Big[\frac{\frac{1}{n}-\zij}{k\lambda} + 
  \log\frac{1+\frac{\zij}{k\lambda}}{\left(1+\frac{1}{k\lambda}\right)^{\frac{1}{n}}}\Big]\Big\}\Big].
 \end{align*} 
 Now we replace the series by a finite sum, with a controlled error,
 by means of the Euler--MacLaurin formula. 
 Denoting by $g(k)$ the expression in the internal square brackets (which depends also on $\zij$),
 
\begin{equation*}
\sum_{k=m}^\infty g(k)=\int_m^\infty g(x)dx + \frac{1}{2}g(m)-\sum_{r=1}^s\frac{B_{2r}}{(2r)!}g^{(2r-1}(m)+R_s,
 \end{equation*}
 with the remainder term satisfying
 \begin{equation*}
|R_s|\leq\frac{|B_{2s+2}|}{(2s+2)!}|g^{(2s+1}(m)| \ ,
 \end{equation*}
and where $B_r$ are the Bernoulli numbers, that can be defined recursively as
\begin{equation*}
B_r=-\Sum_{k=0}^{r-1}\frac{n!B_k}{k!(r+1-k)!}\ ,\quad B_0=1\ .
 \end{equation*}
  
 The formula is true under the conditions
\begin{enumerate}
  \item[(i)]
  $g^{(2s+2}(x)g^{(2s+4}(x)>0\ ,\quad\text{for $x\in[m,\infty]$}$\ ,
  \item[(ii)] 
  $\lim_{x\to\infty}g^{(2s+1}(x)=0$\ .
\end{enumerate} 

If we call $\bar f$
the approximation of the estimated density $\hatf$ when disregarding the 
remainder $R_s$,
 and $M:=\exp\{\max_{i,j} |R_s|\}$, then 
  \begin{equation*}
   \bar f M^{-n}\leq\hatf\leq \bar f M^n \ .
  \end{equation*}
Thus to obtain a final relative error $\eta$, we have to find an $\varepsilon$ such that 
$\varepsilon\ge\max_{i,j}|R_s|$ and 
$\exp\{n\varepsilon\}\leq(1+\eta)$. This amounts to take 
\begin{equation*}
  \varepsilon=\frac{1}{n}\log(1+\eta)\ ,
\end{equation*}
and to find natural numbers $s$ and $m$ such that 
$\max_{i,j} |R_s|<\varepsilon$, and satisfying the conditions of the Euler-MacLaurin formula.
In this way, we will finally get the approximation
    \begin{align*}
     \bar f(x)=\ \frac{C}{N}\Sum_{i=1}^N\exp\Big\{\sum_{j=1}^n\Big[&
      \frac{1}{\lambda}(-\gamma(\frac{1}{n}-\zij)+\zij\log (x_j))\\
      &+\Sum_{k=1}^{m-1}g(k) +\int_m^\infty g(x)dx + \frac{1}{2}g(m) \\
      &-\Sum_{r=1}^s\frac{B_{2r}}{(2r)!}g^{(2r-1}(m)\Big]\Big\}\ ,\nonumber
    \end{align*}
with 
  \begin{equation*}
    (1+\eta)^{-1}\leq  {\hatf}/{\bar f}\leq(1+\eta)\ .
  \end{equation*}    
  
  The conditions to apply the Euler-MacLaurin formula are in our case always fulfilled for 
  very small integers $m$ and $s$, when taking $\eta=10^{-4}$. The minimal ones are readily 
  found by simple search.

\subsection{Sampling strategy}\label{subsec:metod}

Our real dataset contains many cells in which one or more land covers are not present.
Hence, the theoretical distribution from which they are taken does not actually possess a density
on the simplex $\Delta$. 
However, we can assume the existence of a density on the subsimplices obtained by
restricting some of the coordinates to be zero. Indeed, the resolution of our data
is sufficient to estimate the density on each subsimplex, using the points
that lie on it, except in a few cases.

  If $f_\delta$ is the theoretical density on the subsimplex $\delta$, and $q_\delta:=P\{p\in \delta\}$ is the theoretical
  probability that one random point of $\Delta$ lie on the subsimplex $\delta$, the overall probability distribution
  can be described as 
\begin{align*}
  P\{p\in A\}&=\sum_\delta q_\delta\cdot P\{p\in A\cap \delta \mid p\in \delta\}  
  \\
             &=\sum_\delta q_\delta\cdot \int_{A\cap \delta} f_\delta(x)\,dx
  \ ,
\end{align*}  
  for any Borel set $A\subset \Delta$, and where the sum runs over all subsimplices. 

To estimate the distribution of the whole dataset we can therefore proceed in the following way: 
The probabilities $q_\delta$ can be estimated by the sample proportion $\hat q_\delta$ of points lying in $\delta$;
the densities on each subsimplex $\delta$ can be estimated and approximated as $\bar f_\delta$ by the method just described 
on Subsection 
\ref{subsec:numapprox}. One obtains the estimate
\begin{align*}
P\{p\in A\}&\approx \sum_\delta \hat q_\delta\cdot \int_{A\cap \delta} \bar f_\delta(x)\,dx
\ .
\end{align*}

Although there is no a explicit form for the densities $\bar f_\delta$, we can evaluate 
them at arbitrary points $x$ and apply the acceptance/rejection
method to simulate a large sample following this distribution. Specifically:
\begin{enumerate}
  \item 
  Choose randomly a subsimplex $\delta$ with probability $\hat{q}_\delta$. \label{choose_d}
  \item 
  Generate a random vector $x$ with uniform distribution on $\delta$, with the method of 
  Section \ref{sec:unifdist}.  
  \label{generar_punt}
  \item 
  Generate a random number $u$ with uniform distribution on $[0,1]$ and evaluate
\begin{equation*}
uC_\delta\leq \bar f_\delta(x) \ .
\end{equation*}
  If the inequality holds true, accept $x$ as a new point of the sample;
  otherwise, reject it and go back to step \ref{generar_punt}.
  \item 
  Go back to step \ref{choose_d} until the desired sample size is reached.
\end{enumerate} 
In step 3, $C_\delta$ is any constant satisfying 
$
C_\delta\ge\max\{\bar f_\delta(x)\}
$. 
 Ideally, this constant must be an upper bound as tight as possible of the density function
$\bar f_\delta$, in order not to reject too many generated points.
However, we only know this density in a big, but finite, number of points. If, during the
run of the acceptance/rejection method, a value of $\bar f_\delta$ greater than 
the chosen $C_\delta$ is found, then some of the already accepted points must have been 
actually rejected. From the practical point of view, we have
preferred in our case study to take a safe upper bound, so that none of the accepted points
have to be discarded later, despite the larger running times incurred.

The absolute error in the probability of accepting a point $x$ based in the 
approximate density $\bar f$ in step \ref{generar_punt} above, when it would have been rejected
if $\hat f$ could be used, it is bounded by the constant $\eta$. Indeed, the difference
in the probabilities to accept the point in the two cases is 
\begin{align*} 
0&\le\frac{1}{C_\delta}\big(\bar f(x)-\hat f(x)\big)
\le
\frac{1}{C_\delta}\big(\bar f(x)-\bar f(x)(1+\eta)^{-1}\big)
\\
&=
\frac{\bar f(x)}{C_\delta}\big(1-(1+\eta)^{-1}\big)
\le 1-\frac{1}{1+\eta}\le \eta
\ .
\end{align*}
Analogously, one can show that the difference in the probability of rejecting a point 
is less than the same constant $\eta$.

\subsection{Choosing the bandwidth parameter}\label{subsec:lambdafit}

  As mentioned before (see Subsection \ref{subsec:distp}) the goodness of the estimation of
  a density by a kernel method depends heavily on the choice of the bandwidth (or smoothing) parameter
  $\lambda$. In general, the larger the sample size, the smaller the bandwidth should be, or, in other
  words, the less influence each sample point must have on the final estimation.
  
  In our case, the initial sample size is $N=3360$. Although we have to work independently
  on each subsimplex, the bandwidths will tend to be small anyway, as this is what creates the
  numerical problem that we have addressed in Section \ref{subsec:numapprox}.

  In the frequently cited paper by \cite{habbema1974stepwise}, 
  and in \cite{aitchison1985},
  the authors propose to choose the smoothing parameter
  $\lambda$ that maximises the pseudo-likelihood 
  \begin{equation*}
\prod_{i=1}^N\frac{1}{N-1}\sum_{j\neq i}K(x_i, x_j,\lambda I)\ ,
\end{equation*}
where $x$ are the sample points, $N$ is the sample size and $I$ is the identity matrix.

Instead, we will adjust $\lambda$ according to the use that we will make of
the estimated density.  
Namely, we want to approximate the function that maps appropriation levels to
the conditional expectation of the Shannon index given that level:
\begin{equation}\label{eq:funcphi}
a\xmapsto{\phi} \E[H\mid A=a]\ .
\end{equation}

To this end, we proceed
with the following steps, on each subsimplex: 
\begin{enumerate}
  \item 
  Assume the points in the subsimplex follow a Dirichlet distribution.
\item Estimate 
the parameters $\alpha$ of the distribution (\ref{eq:fDirichlet}). We have used the maximum likelihood method  
implemented in the function \verb+dirichlet.mle+ of the R package 
 \verb+sirt+.
\item Generate a large number of points (e.g. $10^6$) $Y$ with the estimated distribution. 
These data plays the role of 'synthetic population' in this process.
\item Sample a subset $Z$ of $Y$ of the same size as the part of the real sample that lies on the
subsimplex.
These data $Z$ is used as the 'synthetic sample' for the next steps.
\item For a given value of $\lambda$, apply the procedure explained in \ref{subsec:metod} to simulate 
a sample $X_\lambda$ of the estimated density (say, of size $10^4$).\label{step:eachlambda}
\item Measure the fit of the simulated data with the 'synthetic population' $Y$ using the   
integrated square error
\begin{equation}\label{eq:squareerror}
\int_{w_1}^{w_n} \left(\phi_Y(a)-\phi_{X_{\lambda}}(a)\right)^2\, da\ ,
\end{equation}
 where $\phi_Y$ and $\phi_{X_{\lambda}}$ are the functions (\ref{eq:funcphi}) for $\phi$ corresponding
 respectively to the population $Y$, and to the sample $X_\lambda$. \label{step:squareerror}

\item Repeat steps \ref{step:eachlambda}--\ref{step:squareerror} to choose $\lambda$ that minimises (\ref{eq:squareerror}).

\end{enumerate}
Some remarks are in order about the scheme above: 
\begin{enumerate}
  \item [a)]
In our case study, it is possibly not true that the data can be well represented
by a Dirichlet distribution; if we knew it were, then 
we would be better off adopting directly the density that results from the maximum likelihood estimate.
However, we use it at this point as a proxy because of its support on the simplex, and only to obtain
a plausible bandwidth; using the uniform distribution on the subsimplices for the same purpose
will be even more inadequate.
  \item [b)]
The sample sizes of $Y$ and $X_\lambda$ are arbitrary. They should simply look like a (big) population
and an (also big) sample from it. On the contrary, we think that it is realistic to make the size of
$Z$ equal to the size of the real data at hand.
The integral in (\ref{eq:squareerror}) cannot be computed exactly, because the function (\ref{eq:funcphi})
cannot be either. We discretise the values of $A$ to obtain a stepwise approximation of $\phi$, 
so that the integral is in fact approximated by a finite sum. But this is fine, since the final result
will necessarily be given as a discretised function.
  \item [c)]
Finally, the integrated square error is not the only possible criterion for the choice of $\lambda$;
others can be used, depending on the application sought. 
\end{enumerate}

\section{Results}\label{sec:results}

In this section we present the results of the procedures proposed in Sections \ref{sec:unifdist}
and \ref{sec:realdata} when applied to the data of the case study described in the introduction.
All figures referenced have been grouped together at the end of the paper, for easy comparison.

The four types of land covers are: the 
semi-natural land covers, with lowest human intervention (forest, scrubland, prairie and bedrock, 
and wetland), $p_1$; 
the cropland, both irrigated and dry crops, $p_2$; the land covers with groves, $p_3$; and 
the urban and industrial surfaces, $p_4$.

This grouping has been established according to the similarity in the weights of the 
original ten land covers, the latter
taken from \cite{Marull2015b}, and
each type is assigned the mean of the original weights (see Table \ref{tab:w}).
There are different values for each year, 
due to the changes in the exploitation of land covers over time.
From 1956 to 2000 there is a general reduction in the values of $w$. 
 It is known that in the last decades of the twentieth century
there has been in Mallorca a progressive abandonment of the arable land, inducing 
an expansion of forests, from which humans extract little profit \cite{Marull2015b}.

\begin{table}[h]
\centering
\begin{tabular}{|c|cccc|}\hline
\rule{0pt}{2ex}
 year&$w_1$&$w_2$&$w_3$&$w_4$\\\hline
 \rule{0pt}{2.5ex}
 1956\ &\ 51.042&\  78.880&\  89.993&\  95.730\\
 1973\ &\ 43.958&\  76.200&\  85.322&\  94.792\\
 2000\ &\ 48.542&\  74.978\ & 81.837&\  93.958\\\hline
\end{tabular}
\caption{$w$ values for each year.}
\label{tab:w}
\end{table}

The real data is distributed in subsimplices as described in Table \ref{tab:subsimplex}. 
As we can see there, the dominant subsimplex in 
1956 and 1973 is the one comprising 'semi-natural', 'cropland' and 'groves' 
covers. Such combinations are usually referred as \emph{mosaic landscapes}.
Their frequency clearly declines in 2000, where the combination of 'semi-natural' and 'cropland'
prevails.
  
In Figure \ref{fig:real_HHANPP}, a scatter plot of the joint values 
of $H$ and $A$ is shown, for each of the three times periods (1956, 1973 and 2000). 
Recall that the support of the feasible pairs has the irregular greyed shape that we saw in 
Figure \ref{fig_espH_4}, with the `legs' of the region resting over the weight values in the
horizontal axis; hence the white empty zones in the scatter plot. 
Dots are plotted with some degree of transparency; the apparently solid lines describing arcs
between the legs are points whose corresponding proportions $p$ lie in the edge joining 
two vertices of the simplex. Some of these edges are more populated than others, 
or more evenly distributed, and those arcs are therefore 
more noticeable in the figure.

In Figure \ref{fig:sim_HHANPP}, the same scatter plot of the pairs $(A,H)$ 
is depicted, for the enlarged dataset obtained 
by the sampling method of Subsection 
\ref{subsec:metod}, and the three corresponding time periods. 
Table \ref{tab:subsimplex} shows the $\lambda$ values on each 
subsimplex obtained following the optimisation procedure of Subsection \ref{subsec:lambdafit}.
Of course, vertices of the simplex does not have a density. Also, we have not estimated a density for 
subsimplices with less than 30 data points; instead, we have sampled them as a discrete equally
probable population. The threshold of 30 is arbitrary. 
 
\begin{figure}[h!]
\captionsetup{width=0.4\textwidth}
\begin{minipage}{0.45\textwidth}
  \centering
  \begin{subfigure}[h]{1\columnwidth}
    \centering
    \includegraphics[width=0.9\linewidth]{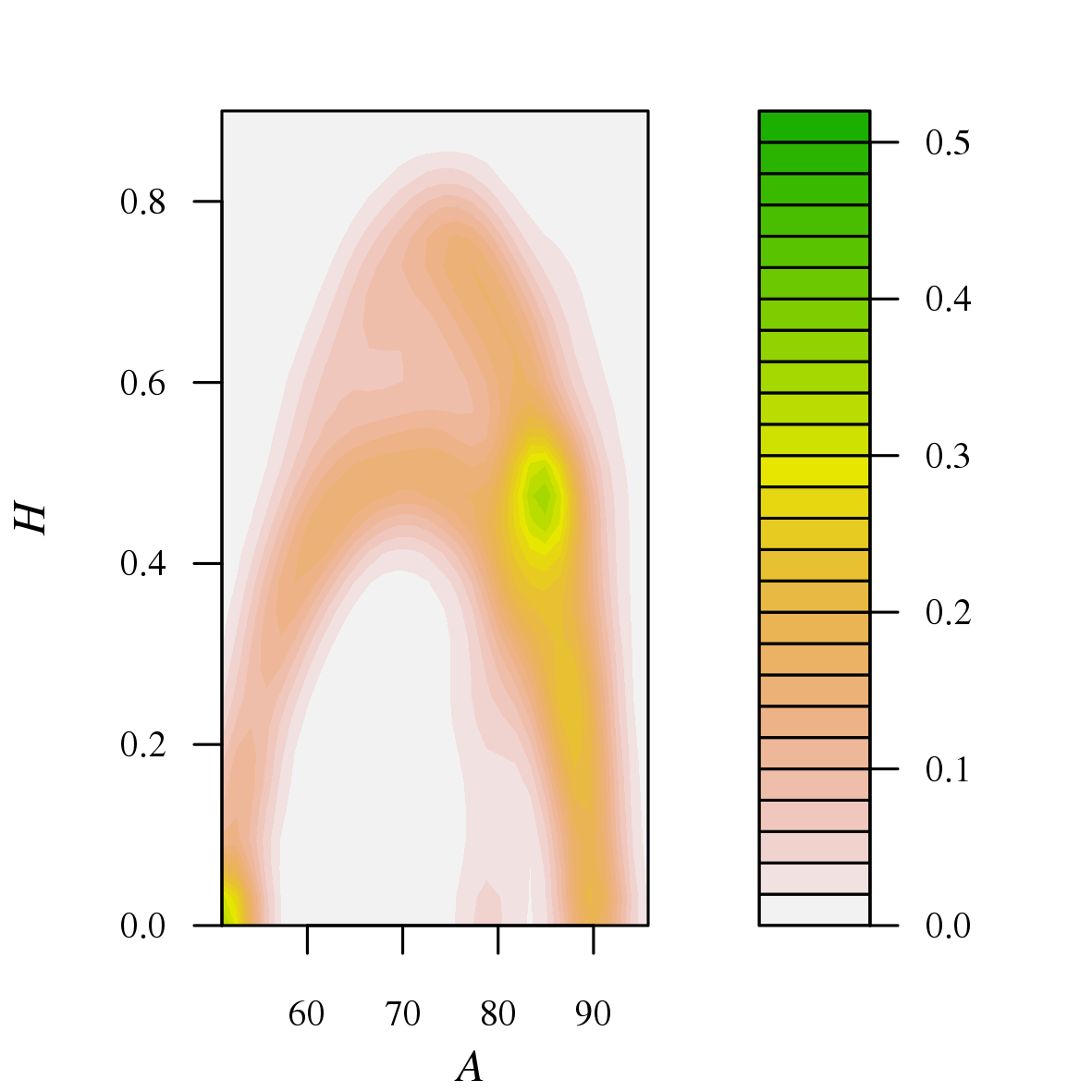}
    \caption{1956}
  \end{subfigure}\\
  \begin{subfigure}[h]{1\columnwidth}
    \centering
    \includegraphics[width=0.9\linewidth]{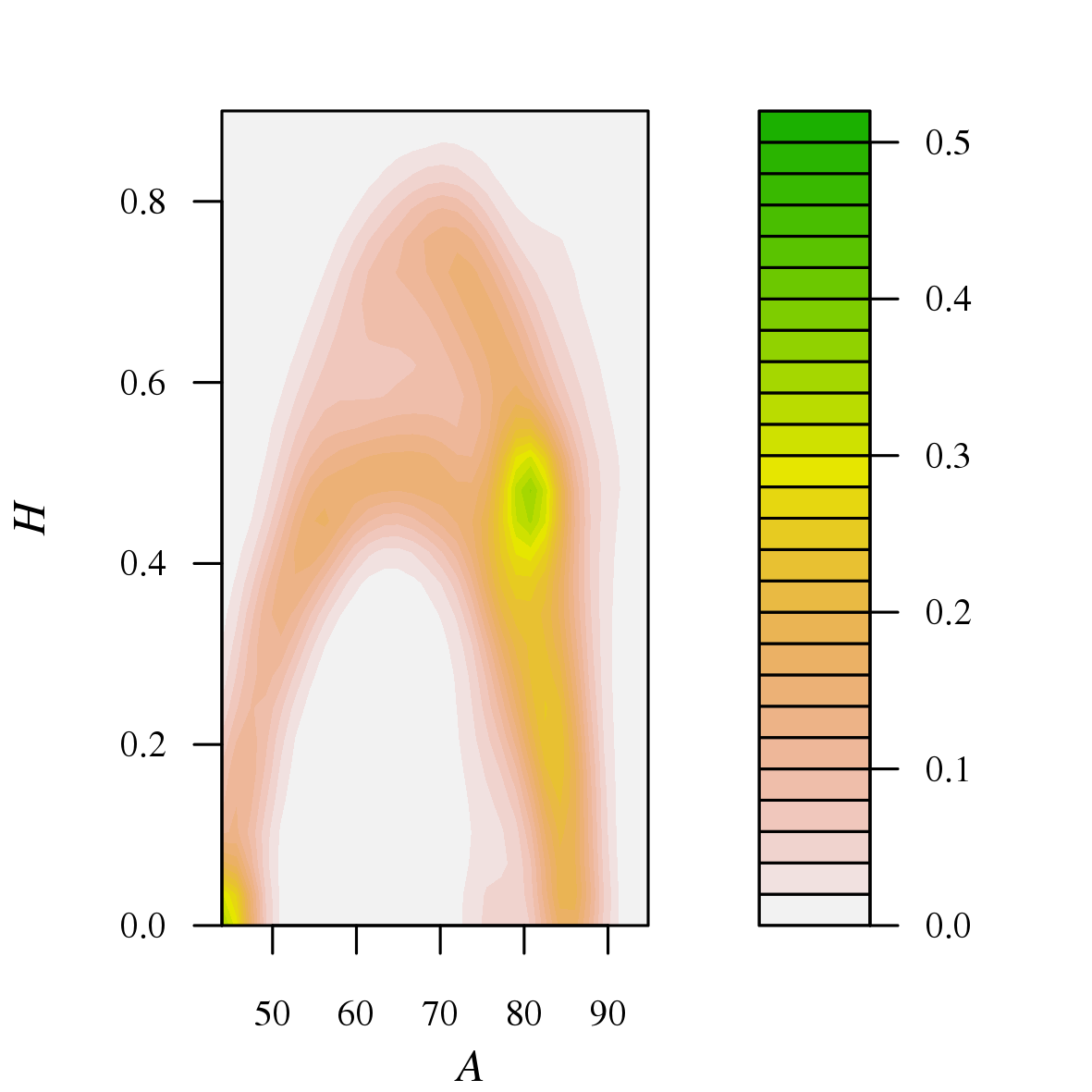}
    \caption{1973}
  \end{subfigure}\\
  \begin{subfigure}[h]{1\columnwidth}
    \centering
    \includegraphics[width=0.9\linewidth]{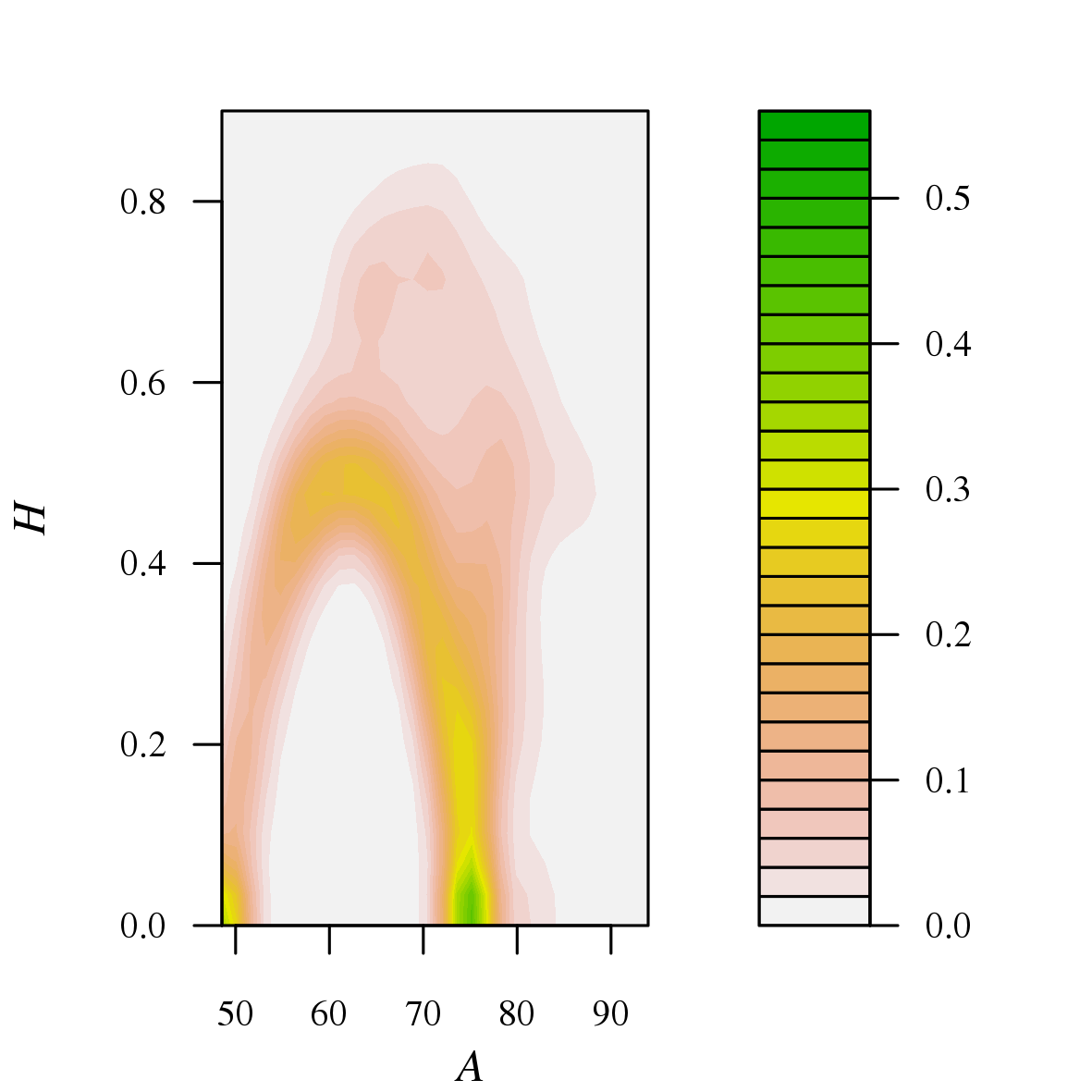}
    \caption{2000}
  \end{subfigure}
  \caption{`Filled-contour plot' of the two-dimensional of $(A,H)$, estimated from real data.}
  \label{fig:rden_HHANPP}
  \end{minipage}
\begin{minipage}{0.45\textwidth}
  \begin{subfigure}[h]{\columnwidth}
    \centering
    \includegraphics[width=0.9\linewidth]{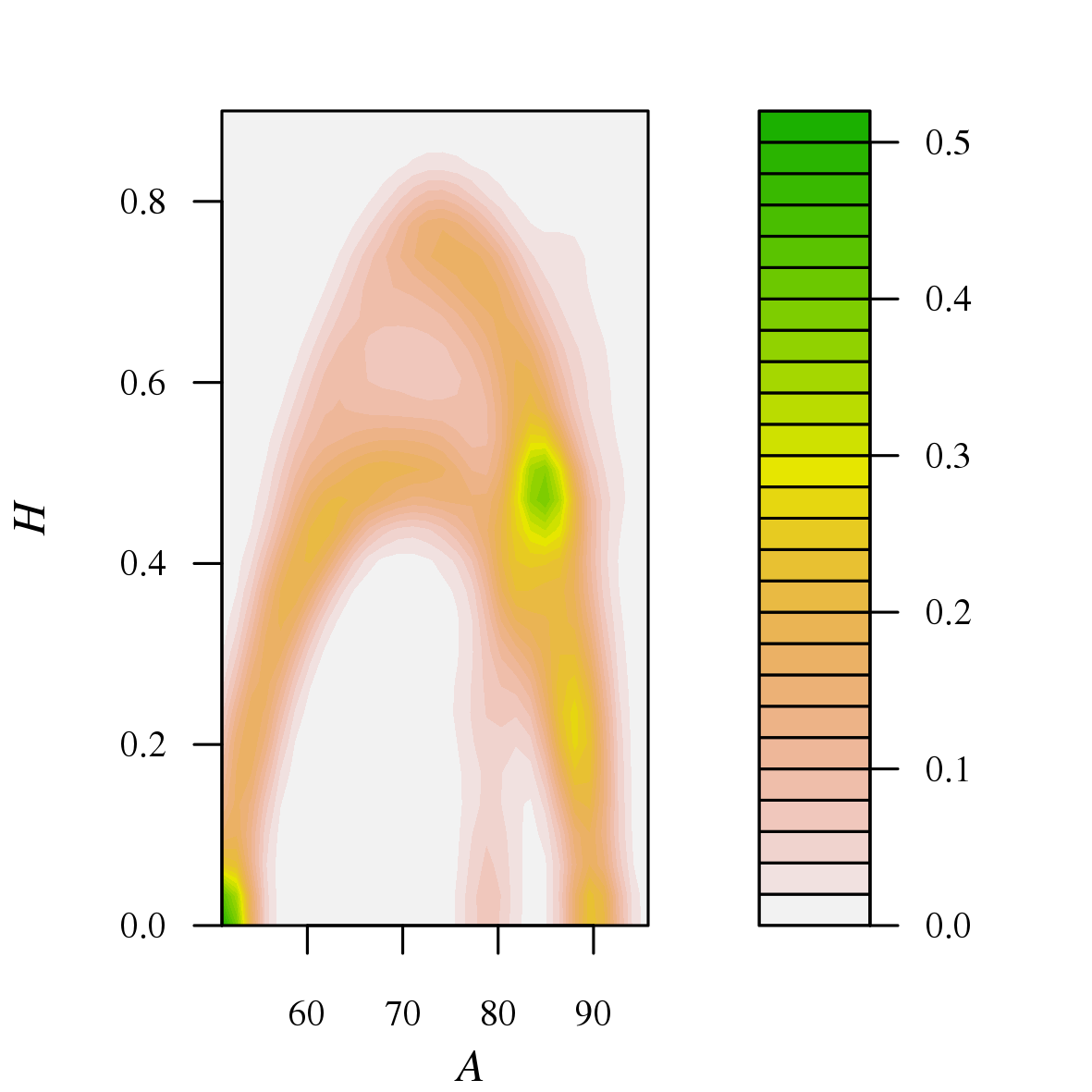}
    \caption{1956}
  \end{subfigure}\\
  \begin{subfigure}[h]{\columnwidth}
    \centering
    \includegraphics[width=0.9\linewidth]{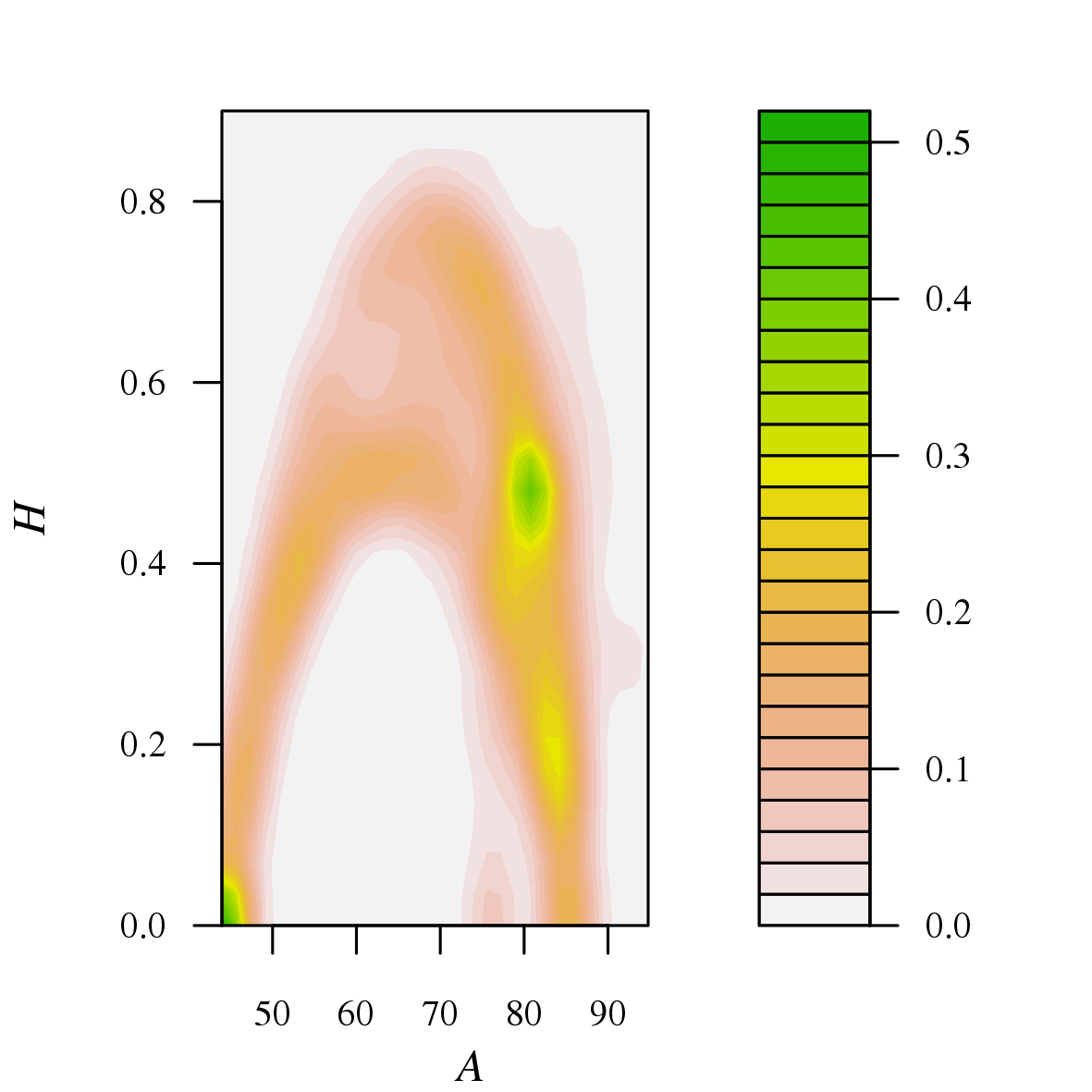}
    \caption{1973}
  \end{subfigure}\\
  \begin{subfigure}[h]{\columnwidth}
    \centering
    \includegraphics[width=0.9\linewidth]{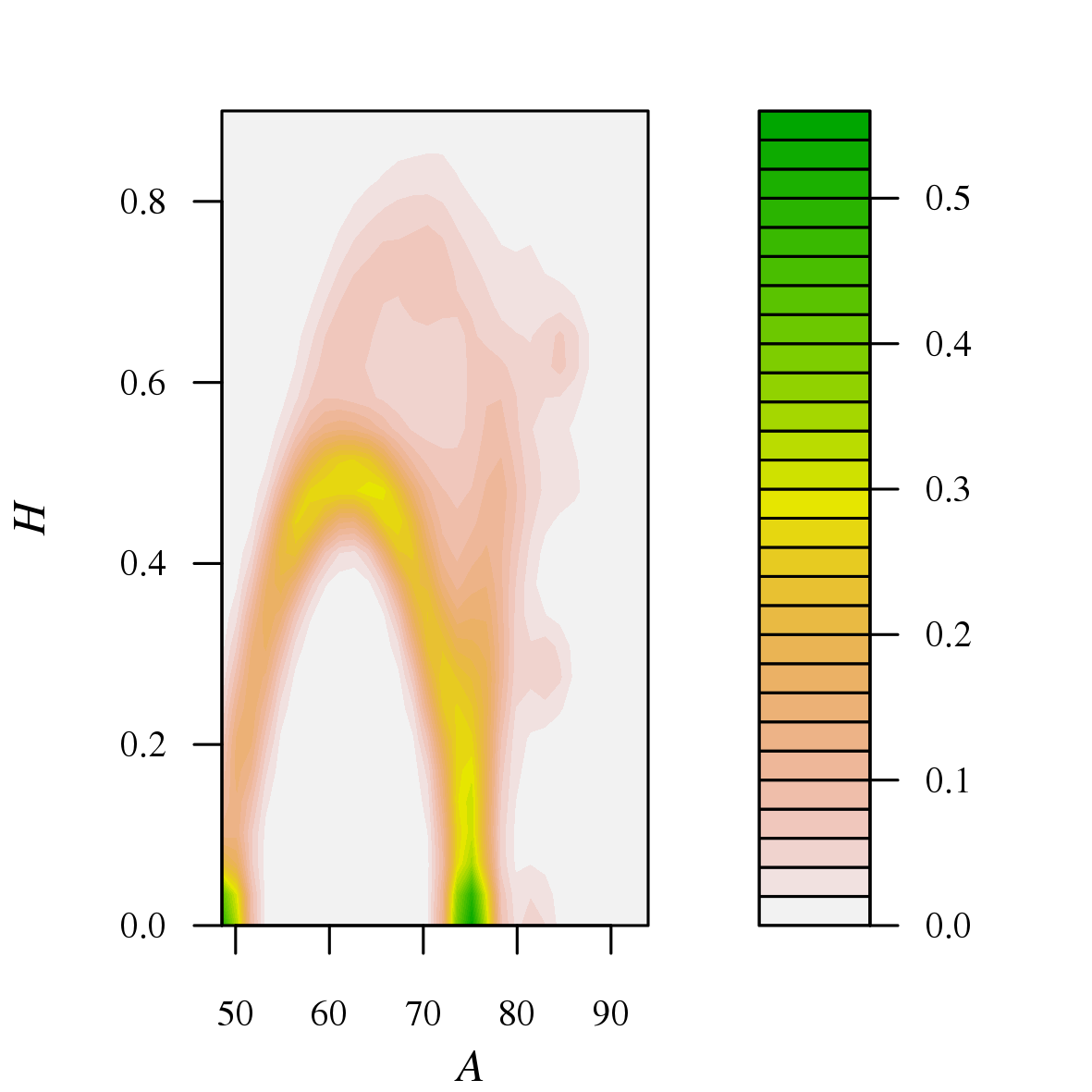}
    \caption{2000}
  \end{subfigure}
  \caption{Analogue of Figure \ref{fig:rden_HHANPP} for the simulated data.}
  \label{fig:sden_HHANPP}
  \end{minipage}
\end{figure}

Except for the number of points, Figures \ref{fig:real_HHANPP} and \ref{fig:sim_HHANPP} look indeed quite similar,
which speaks in favour of our method of estimation of the probability
distribution of the proportions in the simplex. To reinforce this impression,  
in Figures \ref{fig:rden_HHANPP} and \ref{fig:sden_HHANPP}
we compare estimations of the join density of $A$ and $H$ both from the initial data and
for the enlarged sample.
In these figures we have used a simple Gaussian kernel density estimation in the plane,
just to have a visual quick idea of the similarities between the large synthetic sample and the original one,
in order to validate the whole computation of the conditional expectations in Section \ref{sec:realdata}.

\begin{table}[h]
\centering
 {\footnotesize
\begin{tabular}{|c|p{0.42cm}p{0.6cm}|p{0.42cm}p{0.6cm}|p{0.42cm}p{0.6cm}|}\hline
\rule{0pt}{2.5ex}
\multirow{2}{*}{Subsimplices $\delta$}
&\multicolumn{2}{c|}{1956}&\multicolumn{2}{c|}{1973}& \multicolumn{2}{c|}{2000}\\
					&$\ \ \ N_\delta$		&$\lambda$	&$\ \ \ N_\delta$	&	$\lambda$&	$\ \ \ N_\delta$	&	$\lambda$\\\hline
\rule{0pt}{2.5ex}
	$1\quad	0\quad	0\quad	0$	&$\ \ \ 228$		&	-	&$\ \ \ 224$		&	-	&$\ \ \ 226$		&	-	\\
	$0\quad	1\quad	0\quad	0$	&$\ \ \ \ \ 30$		&	-	&$\ \ \ \ \ 27$		&	-	&$\ \ \ 240$		&	-	\\
	$1\quad	1\quad	0\quad	0$	&$\ \ \ 109$		&	0.007	&$\ \ \ \ \ 98$		&	0.029	&$1094$			&	0.001	\\
	$0\quad	0\quad	1\quad	0$	&$\ \ \ \ \ 84$		&	-	&$\ \ \ \ \ 78$		&	-	&$\ \ \ \ \ 24$		&	-	\\
	$1\quad	0\quad	1\quad	0$	&$\ \ \ 787$		&	0.003	&$\ \ \ 766$		&	0.002	&$\ \ \ 199$		&	0.039	\\
	$0\quad	1\quad	1\quad	0$	&$\ \ \ 489$		&	0.013	&$\ \ \ 454$		&	0.026	&$\ \ \ 212$		&	0.009	\\
	$1\quad	1\quad	1\quad	0$	&$   1311$		&	0.006	&$1208$			&	0.006	&$\ \ \ 532$		&	0.004	\\
	$0\quad	0\quad	0\quad	1$	&$\ \ \ \ \ \ \ \ 1$	&	-	&$\ \ \ \ \ \ \ \ 3$	&	-	&$\ \ \ \ \ \ \ \ 7$	&	-	\\
	$1\quad	0\quad	0\quad	1$	&$\ \ \ \ \ \ \ \ 3$	&	-	&$\ \ \ \ \ 12$		&	-	&$\ \ \ \ \ 28$		&	-	\\
	$0\quad	1\quad	0\quad	1$	&$\ \ \ \ \ \ \ \ 8$	&	-	&$\ \ \ \ \ 14$		&	-	&$\ \ \ 144$		&	0.008	\\
	$1\quad	1\quad	0\quad	1$	&$\ \ \ \ \ \ \ \ 8$	&	-	&$\ \ \ \ \ 13$		&	-	&$\ \ \ 298$		&	0.007	\\
	$0\quad	0\quad	1\quad	1$	&$\ \ \ \ \ \ 39$	&	0.027	&$\ \ \ \ \ 51$		&	0.05	&$\ \ \ \ \ 24$		&	-	\\
	$1\quad	0\quad	1\quad	1$	&$\ \ \ \ \ \ 59$	&	0.032	&$\ \ \ 111$		&	0.015	&$\ \ \ \ \ 29$		&	-	\\
	$0\quad	1\quad	1\quad	1$	&$\ \ \ \ 105$		&	0.014	&$\ \ \ 141$		&	0.011	&$\ \ \ 136$		&	0.015	\\
	$1\quad	1\quad	1\quad	1$	&$\ \ \ \ \ \ 99$	&	0.035	&$\ \ \ 160$		&	0.03	&$\ \ \ 167$		&	0.031	\\\hline

\end{tabular}
}
\caption{Subsimplex typologies $\delta$, corresponding to different combinations of land covers 
  (1 indicates presence, 0 absence);
  size $N_\delta$ of each subsimplex, and chosen values of $\lambda$.}
\label{tab:subsimplex}
\end{table}

\subsection{Shannon index conditioned to the appropriation}
In Figure \ref{H_esp} we can see superimposed the plots of $a\mapsto \E[H\mid A=a]$, 
with the assumptions of both
Sections \ref{sec:unifdist} and \ref{sec:realdata}.
The red curve is the analytic result obtained assuming a uniform distribution
of covers, whereas the blue points are the ones we have obtained with the real data
of our case study and the procedure of Section \ref{sec:realdata}.
The vertical grey lines 
indicate the values $w$.

Real data produce, for all time periods and for practically all values of appropriation, 
an expected value of the Shannon index $H$ lower than with the uniform distribution. 
This was absolutely expected,
because in the real dataset rarely all types of cover appear in a single cell 
(only 99 over 3360 cases, see Table \ref{tab:subsimplex}), 
nor the appearing ones look like evenly distributed. Recall that the Shannon index is 
maximal when all proportions coincide.

Concerning the annual evolution, figures show a strong similarity in the expected $H$ 
for 1956 and 1973, whereas there are noticeable differences in 2000. 
First, there is a high decrease around $a=w_2$, motivated by the intensification of $p_2$.
Secondly, the 
expectation after $w_2$ increases due to the growth of urban areas combined with other 
land covers. The maximum of the expectation, in fact, jumps to the interval $[w_3,w_4]$.

\subsection{Diversity and urban land cover}

At the scale we are working in, one may consider that the urban cover is not a real habitat for living
species (except humans). It has been proposed in \cite{marull2016energy} to use a variation of the Shannon index that penalises 
the presence of 
urban areas, as indicator of habitat diversity:
\begin{equation*}
L:=(1-p_u)\Big(-\sum_{i=1}^n p_i\log_np_i\Big)\ ,
\end{equation*}
where 
$p_i$ are the proportions of non-urban covers, inside the total of non-urban surface, 
and $p_u$ is the  
proportion of urban surface in the cell. With the obvious notation,
\begin{equation*}
p_i=\frac{S_i}{S_{\text{Cell}} - S_u} \ ,\quad p_u=\frac{S_u}{S_{\text{Cell}}}\ .
\end{equation*}
The maximum of $L$ is 1 and it corresponds
to $p_i=\frac{1}{n}$, $p_u=0$, with appropriation $A=\frac{1}{n}\sum_{i=1}^nw_i$.

\begin{figure}[h!]
\captionsetup{width=0.4\textwidth}
\begin{minipage}{0.45\textwidth}
  \centering
  \begin{subfigure}[h]{\columnwidth}
    \centering
    \begin{overpic}[scale=0.69,unit=1mm]{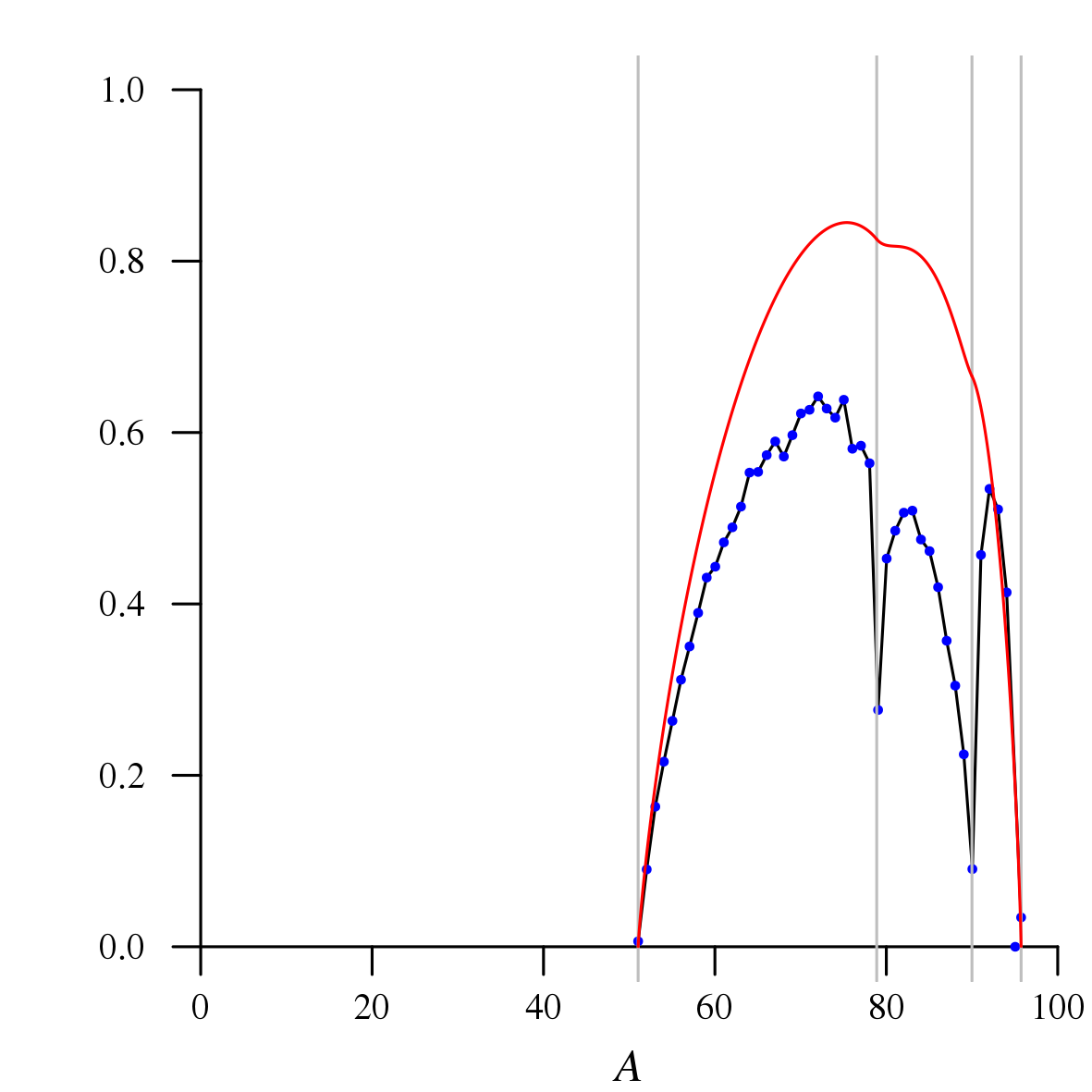}
   \put(-1,-1){\begin{turn}{90}
   {\parbox{1\linewidth}{
      \begin{equation*}
      \E[ H \mid A]
      \end{equation*}}}
      \end{turn}}
 \end{overpic}
    \caption{1956}
  \end{subfigure}
  \begin{subfigure}[h]{\columnwidth}
    \centering
    \begin{overpic}[scale=0.69,unit=1mm]{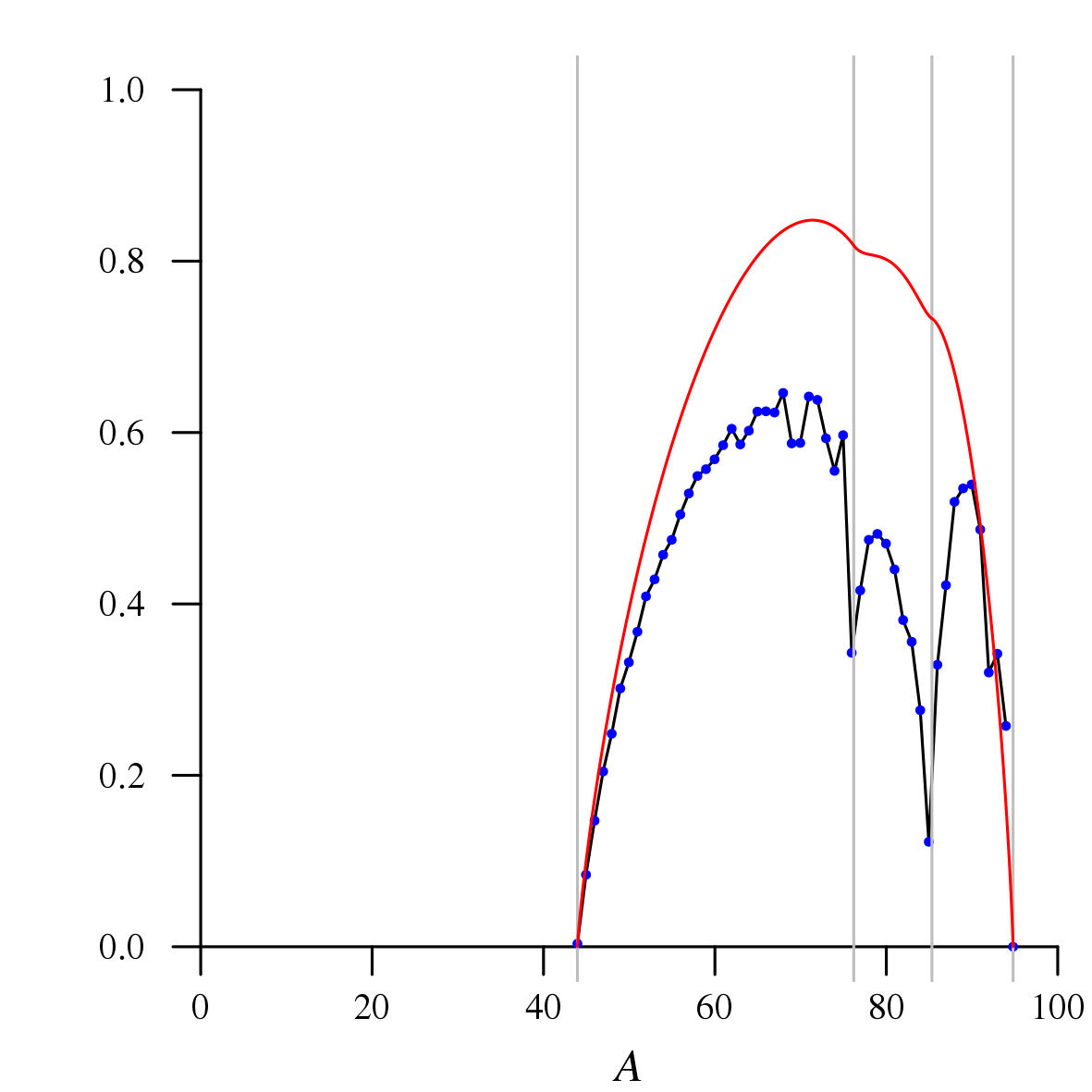}
   \put(-1,-1){\begin{turn}{90}
   {\parbox{1\linewidth}{
      \begin{equation*}
      \E[ H \mid A]
      \end{equation*}}}
      \end{turn}}
 \end{overpic}
    \caption{1973}
  \end{subfigure}
  \begin{subfigure}[h]{\columnwidth}
    \centering
    \begin{overpic}[scale=0.69,unit=1mm]{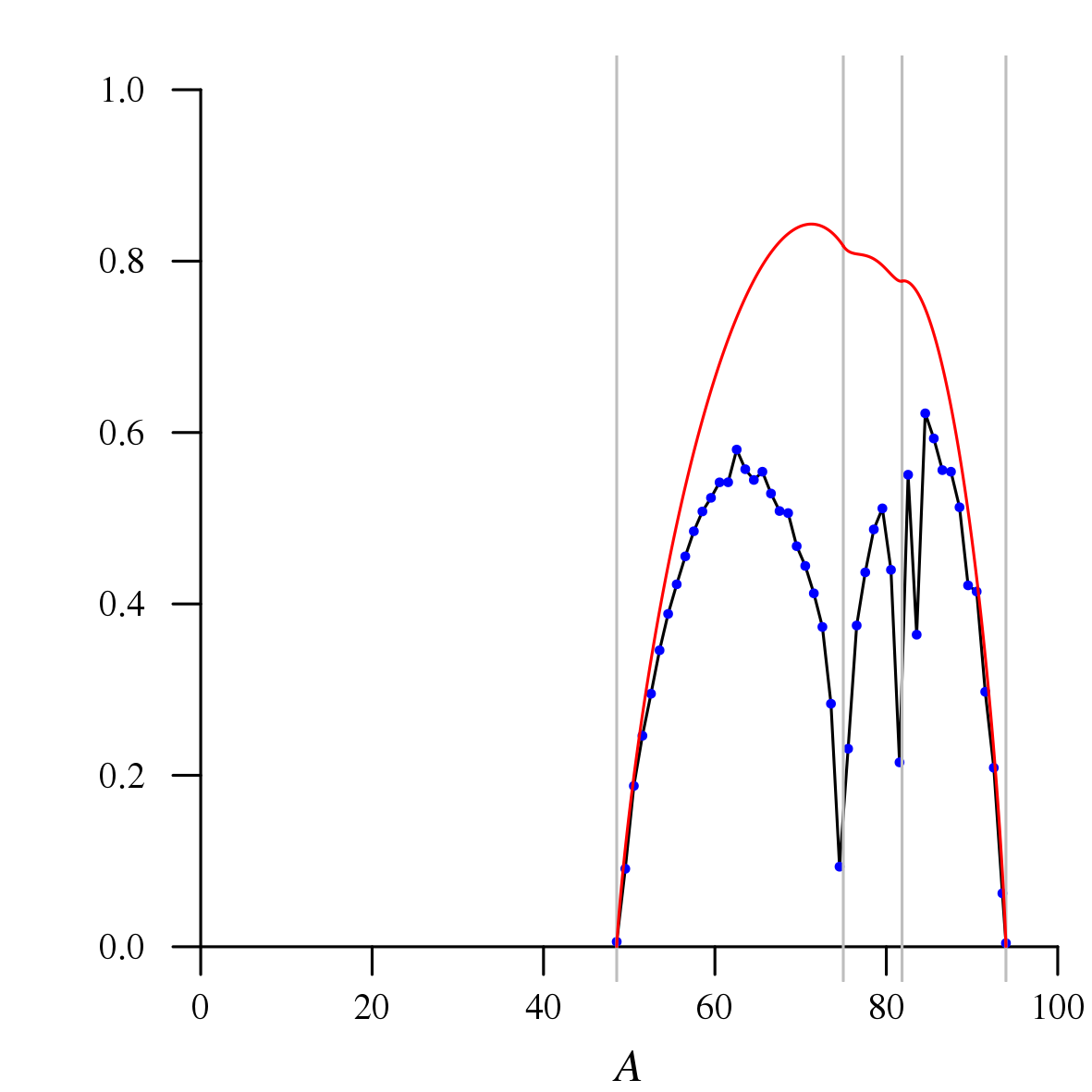}
   \put(-1,-1){\begin{turn}{90}
   {\parbox{1\linewidth}{
      \begin{equation*}
      \E[ H \mid A]
      \end{equation*}}}
      \end{turn}}
 \end{overpic}
    \caption{2000}
  \end{subfigure}
  \caption{Conditional expectation $E[H\mid A=a]$ with the uniform distribution of 
     covers (red curve) and starting form the data of the case study (blue dots).}
  \label{H_esp}
  \end{minipage}
\begin{minipage}{0.45\textwidth}
\begin{subfigure}{\columnwidth}
    \centering
\begin{overpic}[scale=0.69,unit=1mm]{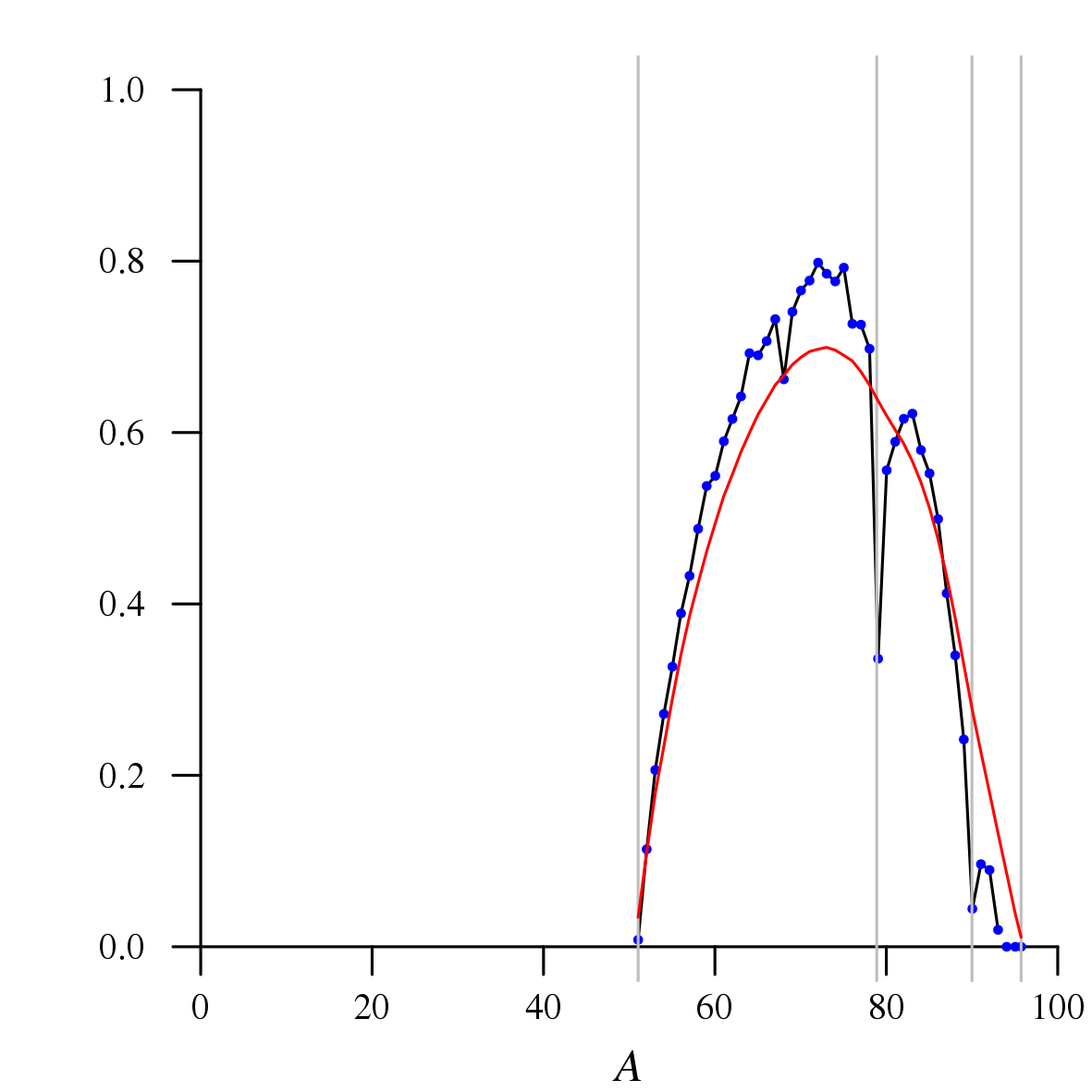}
   \put(-1,-1){\begin{turn}{90}
   {\parbox{1\linewidth}{
      \begin{equation*}
      \E[ L \mid A]
      \end{equation*}}}
      \end{turn}}
 \end{overpic}
\caption{1956}
\end{subfigure}\\
\begin{subfigure}{\columnwidth}
    \centering
\begin{overpic}[scale=0.69,unit=1mm]{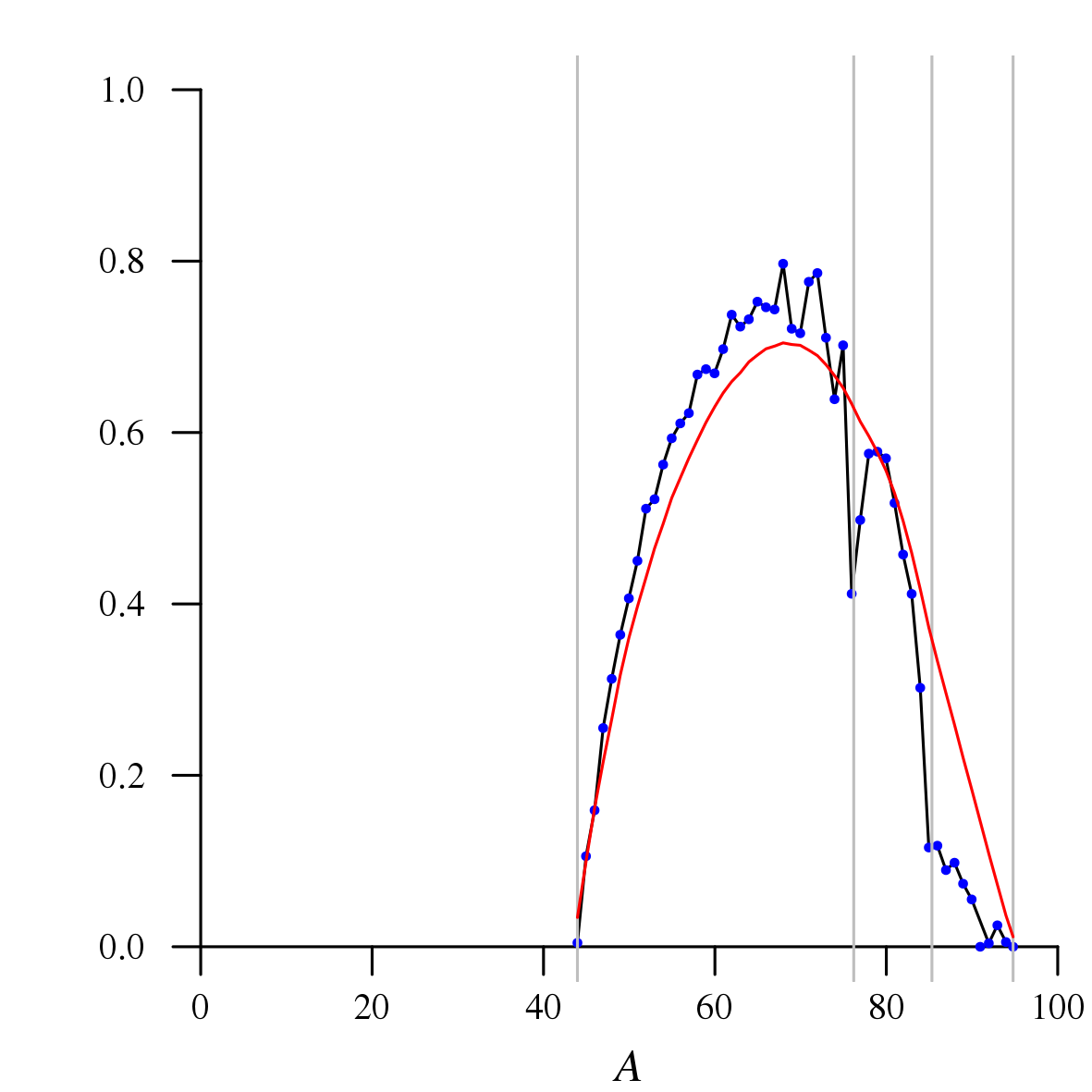}
   \put(-1,-1){\begin{turn}{90}
   {\parbox{1\linewidth}{
      \begin{equation*}
      \E[ L \mid A]
      \end{equation*}}}
      \end{turn}}
 \end{overpic}
\caption{1973}
\end{subfigure}\\
\begin{subfigure}{\columnwidth}
    \centering
\begin{overpic}[scale=0.69,unit=1mm]{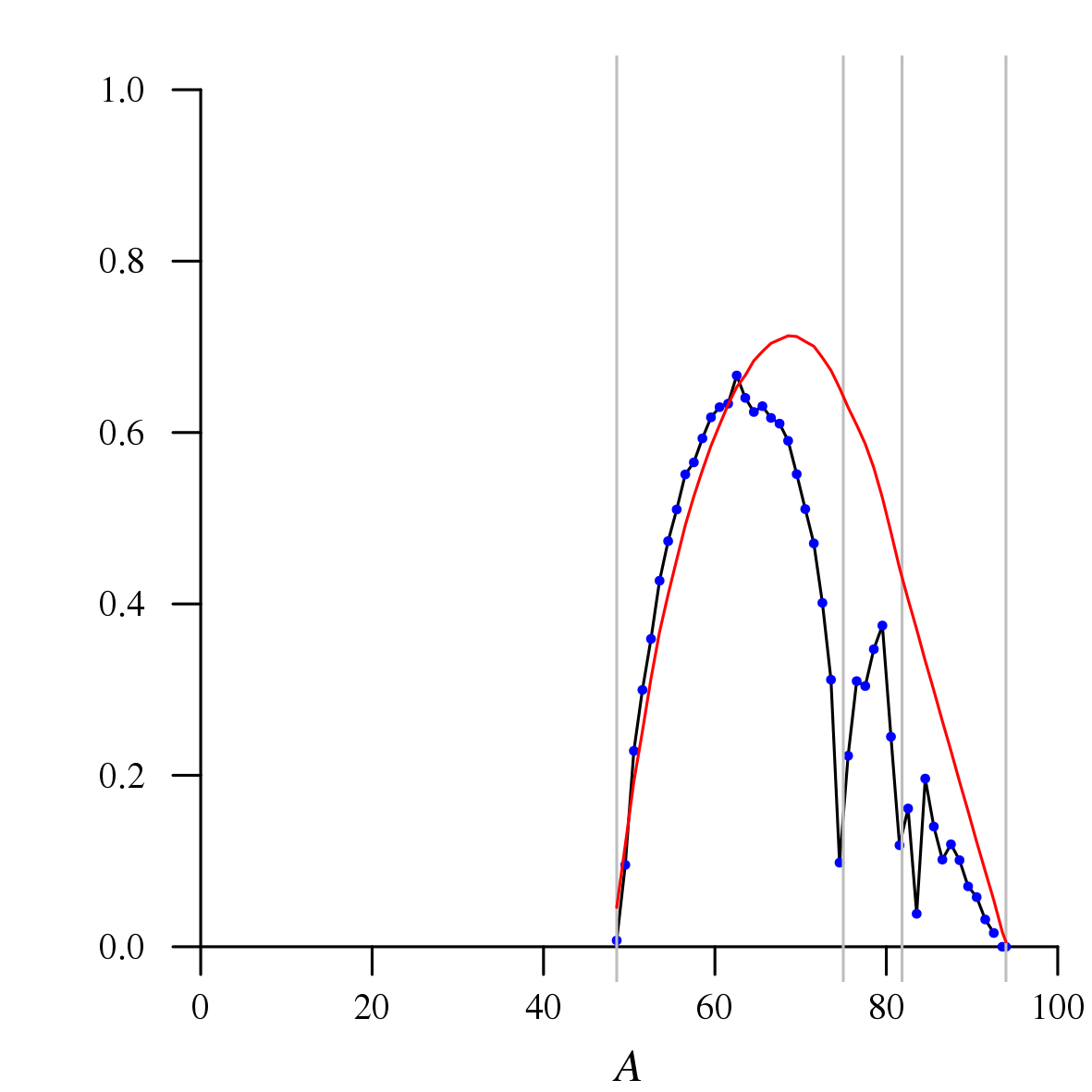}
   \put(-1,-1){\begin{turn}{90}
   {\parbox{1\linewidth}{
      \begin{equation*}
      \E[ L \mid A]
      \end{equation*}}}
      \end{turn}}
 \end{overpic}
\caption{2000}
\end{subfigure}\\
\caption{Analogue of Figure \ref{H_esp} with the indicator $L$.\\\\}
\label{L_esp}
\end{minipage}
\end{figure}

As we pointed out before, 
there is no problem in applying the same methodology of Section \ref{sec:realdata} 
to $L$ or other indices depending only on $p$. 
In Figure \ref{L_esp} one can see the relation we have obtained between the appropriation $A$ and the conditional expectation
$\E[L\mid A]$. Again, the red curve corresponds to the expectation of the index $L$ for each value of $A$
when the land covers, including $p_u$, are uniformly distributed. In contrast with the case of $H$, 
this conditional expectation is in some intervals smaller 
than the values derived from the real data,
represented by the blue 
dots.

The temporal evolution of  $\E[L\mid A]$ in this figure reveals an evident 
change in the landscape structure from 1973 to 2000, essentially due to the urban growth and the 
decline of mosaic prevalence in that
time interval. 
The changes can be partially explained with the help of Table \ref{tab:subsimplex}. 

\begin{figure}[h!]
\captionsetup{width=0.4\textwidth}
\begin{minipage}{0.45\linewidth}
\begin{subfigure}{1\columnwidth}
    \centering
    \includegraphics[width=0.9\linewidth]{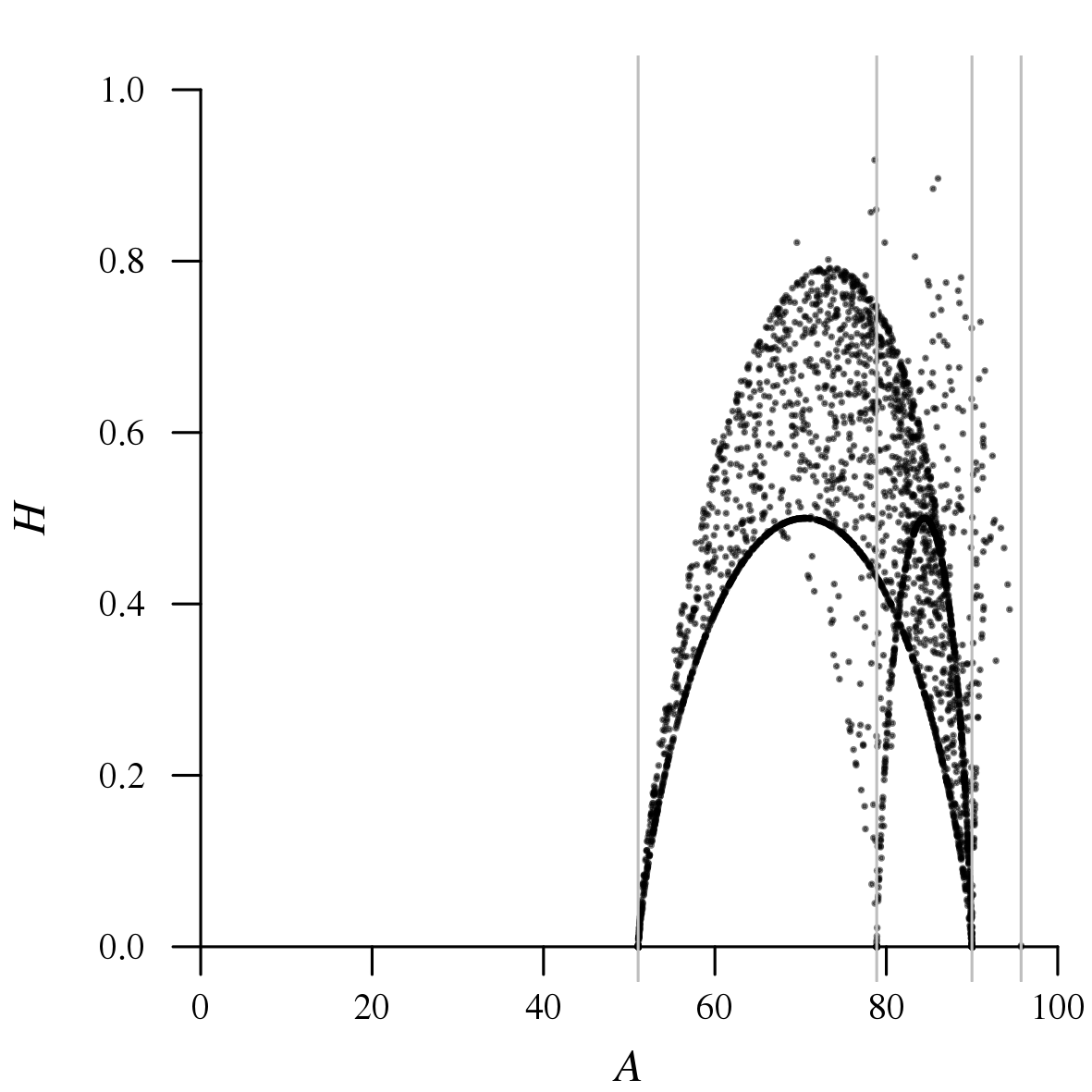}
    \caption{1956}
  \end{subfigure}\\
  \begin{subfigure}{1\columnwidth}
    \centering
    \includegraphics[width=0.9\linewidth]{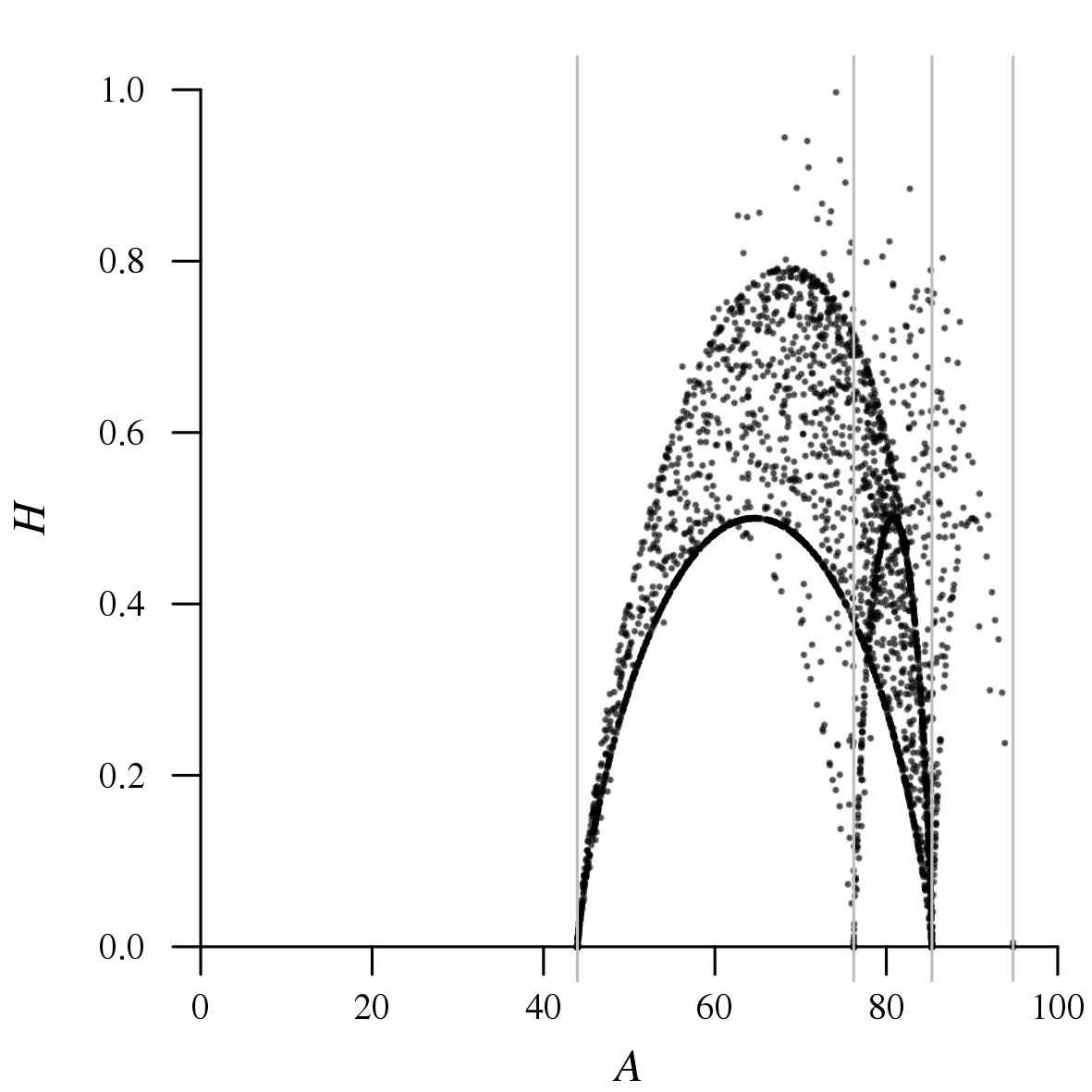}
    \caption{1973}
  \end{subfigure}\\
  \begin{subfigure}{1\columnwidth}
    \centering
    \includegraphics[width=0.9\linewidth]{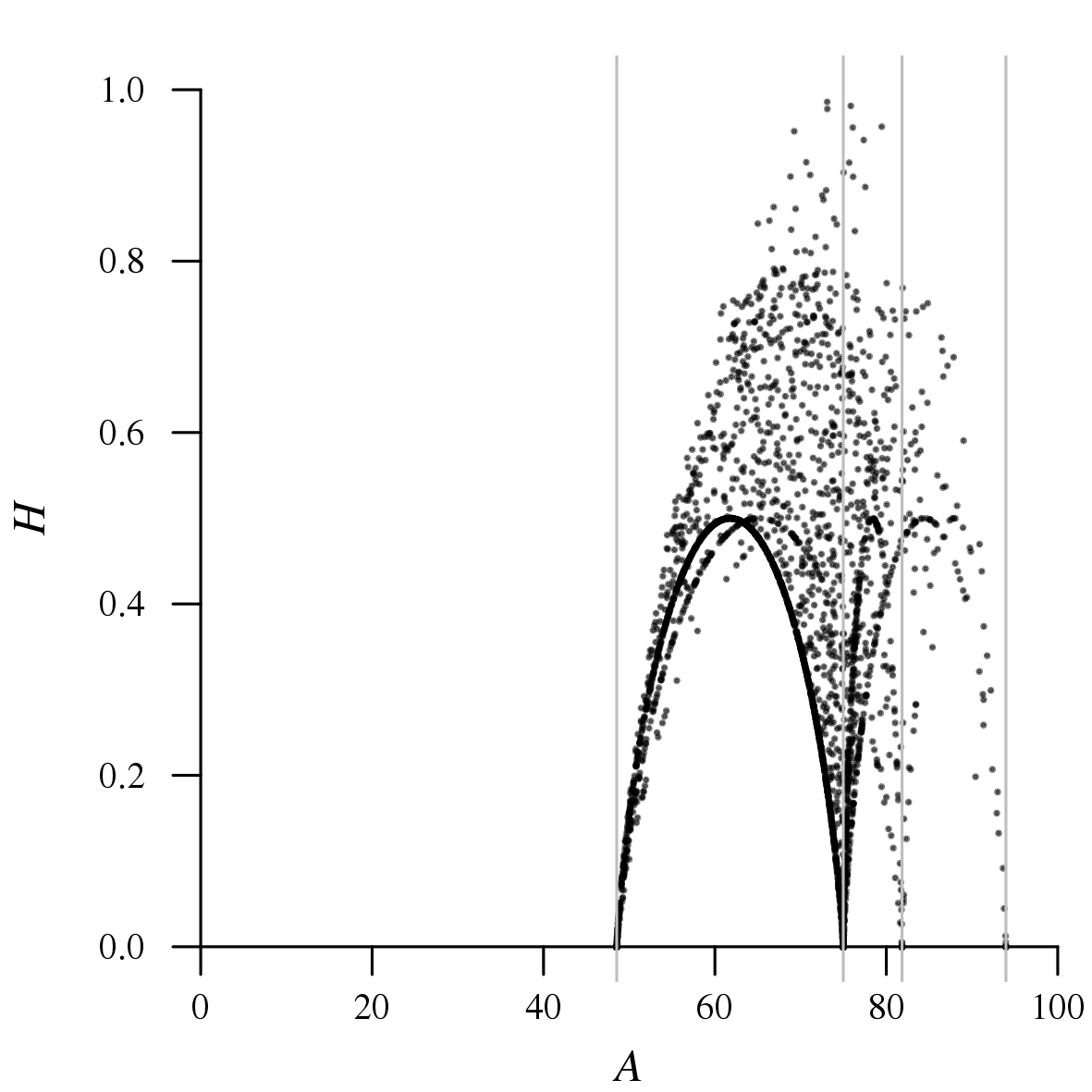}
    \caption{2000}
  \end{subfigure}%
  \caption{Values $(A,H)$ of the real data from Mallorca Island (four land covers, 3360 points).\\}
  \label{fig:real_HHANPP}
  \end{minipage}
\begin{minipage}{0.45\textwidth}
  \begin{subfigure}{1\columnwidth}
    \centering
    \includegraphics[width=0.9\linewidth]{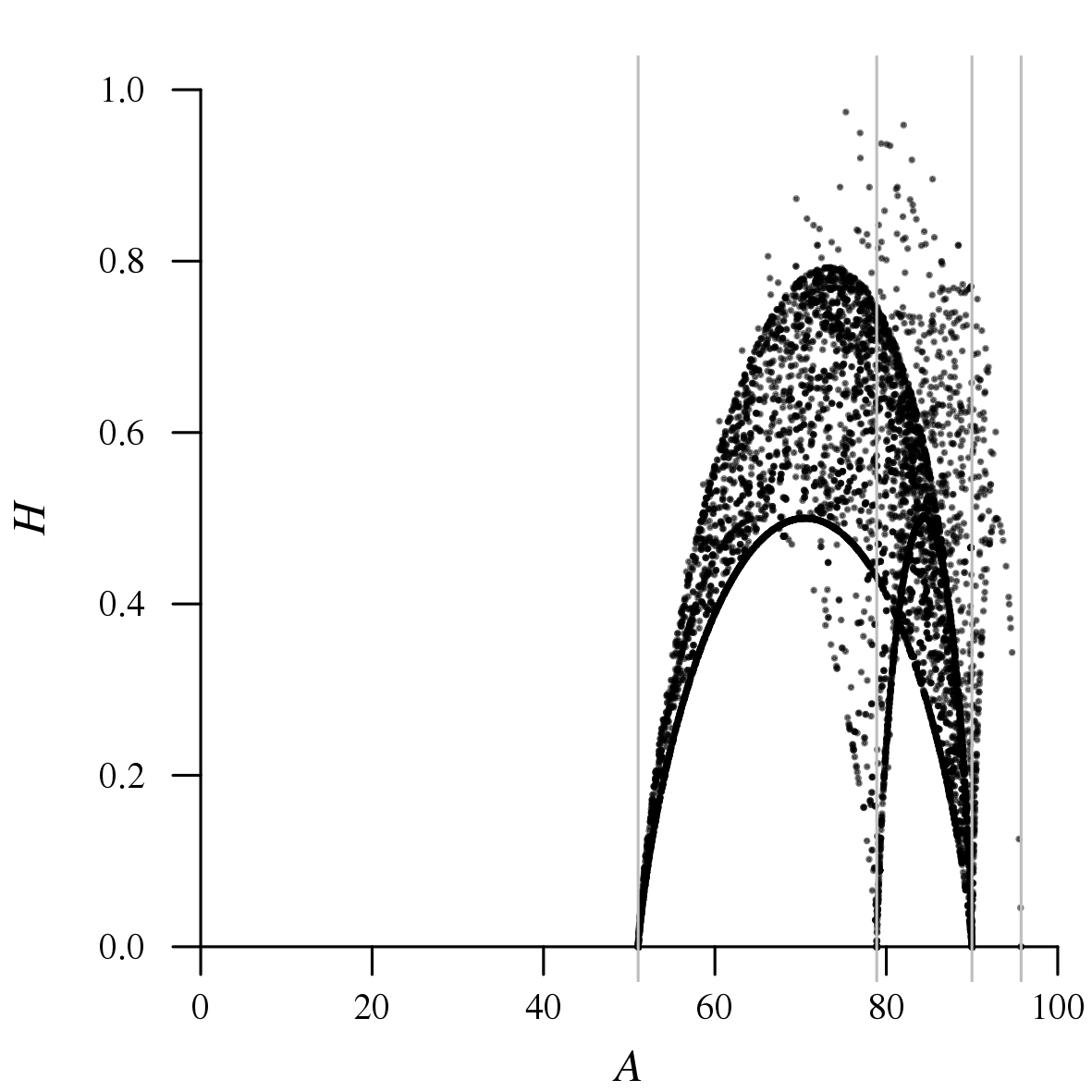}
    \caption{1956}
  \end{subfigure}\\
  \begin{subfigure}{1\columnwidth}
    \centering
    \includegraphics[width=0.9\linewidth]{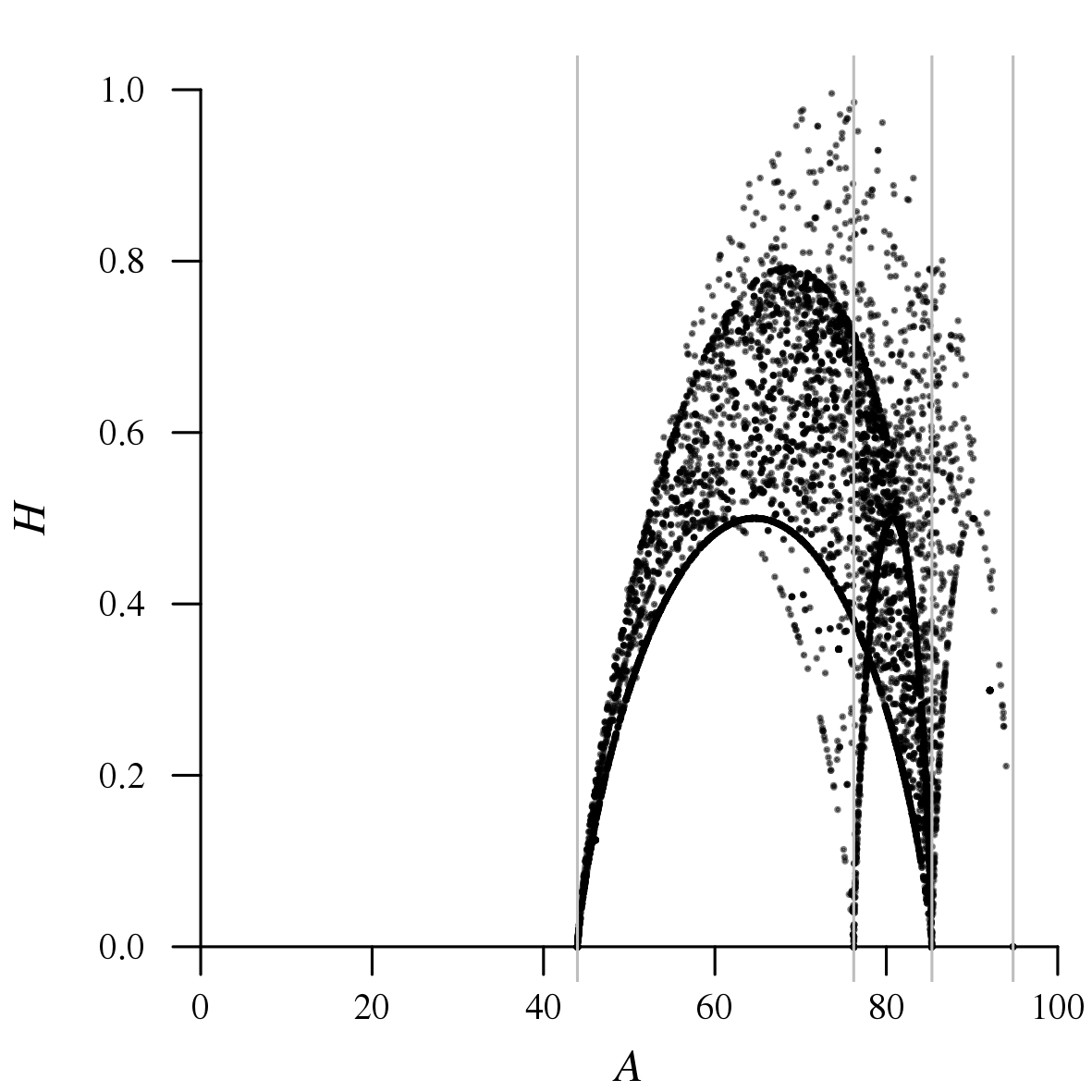}
    \caption{1973}
  \end{subfigure}\\
  \begin{subfigure}{1\columnwidth}
    \centering
    \includegraphics[width=0.9\linewidth]{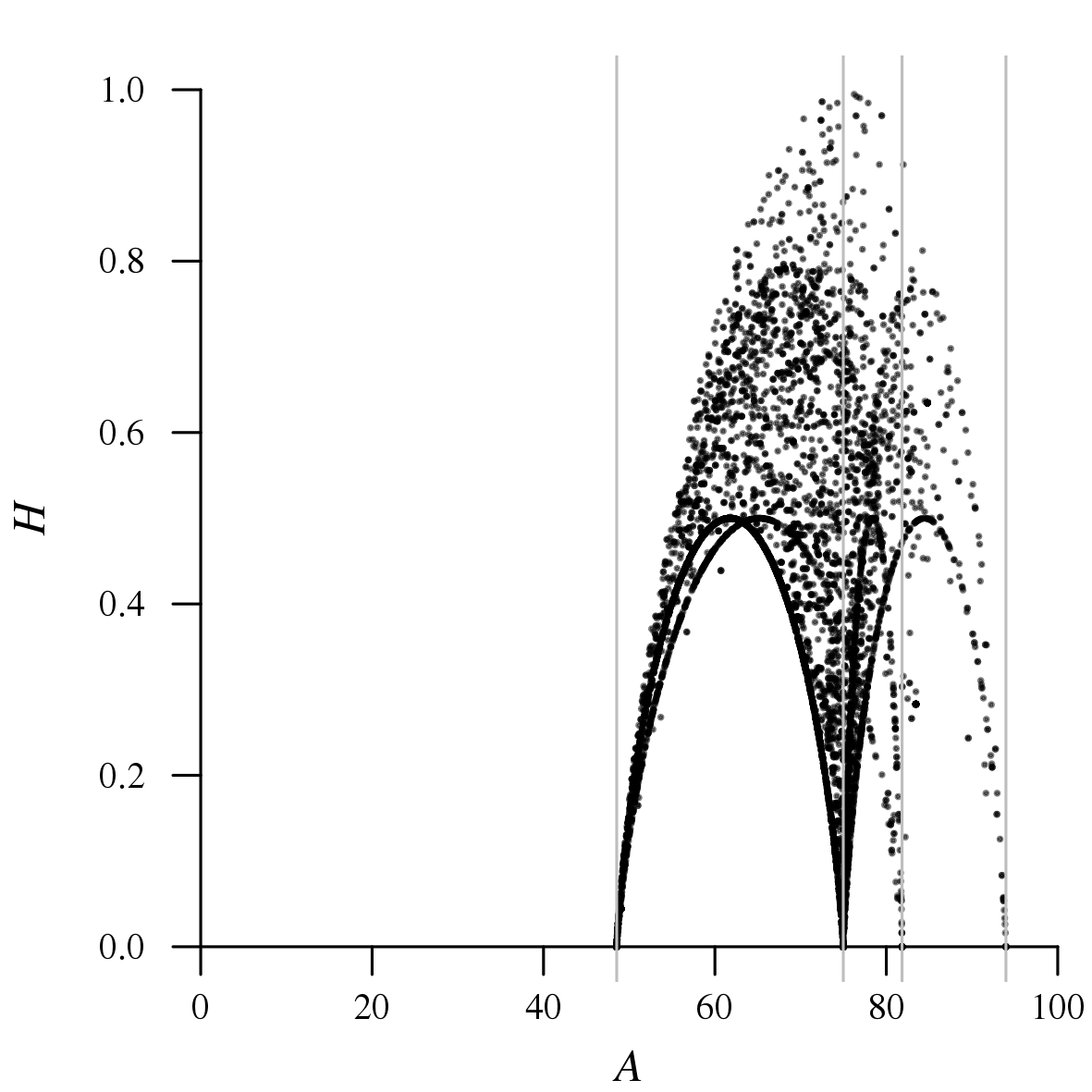}
    \caption{2000}
  \end{subfigure}
  \caption{Simulated values $(A,H)$ generated from the estimated distribution, and a sample size of 
     $10^4$ points.}
  \label{fig:sim_HHANPP}
\end{minipage}
\end{figure}

First, we observe the much lower values around $w_2$: The number of cells where only the second
type of cover is present (0100 in the table), or with the second and the urban cover (0101), have
increased notably, and all of them produce a value $L=0$. Therefore, the mean of the $L$ index 
for values of appropriation in the interval $[w_2,w_4]$ must be lower in 2000 than in 1973. This can
be expected also by comparing the density of points in graphs (b) and (c) of Figure \ref{fig:real_HHANPP} 
or Figure \ref{fig:sim_HHANPP}, around $w_2$.
However,
on $[w_3,w_4]$ this effect is more than compensated by the fact that covers of type $0011$ have decreased; 
without the factor $(1-p_u)$, which is small in this region, the graph will in fact get even higher, as in the case of $\E[H\mid A]$
in Figure \ref{H_esp}. 
The effect of $(1-p_u)$ is almost negligible on $[w_1,w_3]$.

Concerning the interval $[w_1,w_2]$, in which the blue curve tends also to be lower in 2000 than in 
1973 (most notably in the right half of the interval), 
a possible explanation is the smaller number of cells with the non-urban mosaic (1110), together
with the increase of the cells of type (1100), two facts that are of course related. 
Indeed, this produces a higher proportion of values of $L$ at the minimum possible value,
on the arc joining $w_1$ and $w_2$ (the evolution is again apparent on Figures \ref{fig:real_HHANPP} 
and \ref{fig:sim_HHANPP}), and therefore a decrease of the expected value of $L$ given $A\in[w_1,w_2]$.

 While the temporal evolution of the dotted blue curve $\E[H\mid A]$ 
 shows a land cover diversity loss for low values of the appropriation $A$,
 and a gain for high values of $A$,
 the curve based on index $L$ helps to analyse better, in our opinion, the effect of 
 Mallorca urban 
 expansion on habitat diversity. Namely, this effect is reflected in the general decrease of 
 the values of the conditional
 expectation $\E[L\mid A]$ over the whole range of appropriation levels.

\section{Computational notes}

The computations have been done using R with the following setup:
\begin{itemize}\raggedright
\item R version 3.3.1 (2016-06-21), \verb|x86_64-pc-linux-gnu|
\item Base packages: base, datasets, graphics, grDevices, methods, stats, utils
\item Other packages: CDM 5.0-0, knitr 1.13, logspline 2.1.9, mvtnorm 1.0-5,
sirt 1.12-2, TAM 1.995-0
\end{itemize}
The C language has also been used in the most time-consuming routines.

The whole procedure of Section \ref{sec:realdata} is computationally intensive, due to the optimisation step to choose the
right value of the parameter $\lambda$ for each subsimplex, including an acceptance/\-rejection 
simulation for each tentative value. It is not a prohibitive load, though. The computational 
complexity is of course exponential as a function of the number of different covers,
since there are $2^{n+1}-1$ subsimplices in the $n$-dimensional standard simplex $\Delta$ in $\RR^{n+1}$.
For this reason, and to have enough sample size in most of the subsimplices, we grouped
together the ten different covers of the original data into four classes of similar weights.

{\small
\bibliographystyle{plain}
\bibliography{Alabert-Farre-Font_arxiv}   
}

\newpage

\end{document}